\documentclass[pdflatex,sn-basic]{sn-jnl}
\usepackage{graphicx}%
\usepackage{multirow}%
\usepackage{amsmath,amssymb,amsfonts}%
\usepackage{amsthm}%
\usepackage{mathrsfs}%
\usepackage[title]{appendix}%
\usepackage[dvipsnames]{xcolor}%
\usepackage{textcomp}%
\usepackage{manyfoot}%
\usepackage{booktabs}%
\usepackage{algorithm}%
\usepackage{algorithmicx}%
\usepackage{algpseudocode}%
\usepackage{listings}%
\usepackage{natbib}
\usepackage[T1]{fontenc}
\usepackage{siunitx}

\usepackage[version=4]{mhchem}

\usepackage{cleveref} % needs to be last package

\DeclareSIUnit{\yr}{yr}
\DeclareSIUnit{\Myr}{Myr}
\DeclareSIUnit\au{au}
\DeclareSIUnit\micron{\micro m}
\DeclareSIUnit\cm{cm}
\DeclareSIUnit\ppm{ppm}
\DeclareSIUnit\grav{G}
\DeclareSIUnit\bar{bar}
\DeclareSIUnit\wtp{wt\%}

\definecolor{GREEN}{RGB}{0,102,0}

\defcitealias{baumeister2025FundamentalsInterior}{Baumeister, Miozzi, et al. 2025}
\begin{document}

\title{Evolution and observable properties of rocky planet atmospheres}


\author*[1]{\fnm{Marie-Luise} \sur{Steinmeyer}}\email{msteinmeyer@phys.ethz.ch}
\author[2]{\fnm{Lena} \sur{Noack}}%\email{lena.noack@fu-berlin.de}
%\author*[2]{\fnm{Lena} \sur{Noack}}\email{lena.noack@fu-berlin.de}
\author[2]{\fnm{Philipp} \sur{Baumeister}}%\email{}
\author[3,4]{\fnm{Keiko} \sur{Hamano}}%\email{keiko.i.hamano@gmail.com}
%\author[5]{\fnm{Stephanie} \sur{Olson}}\email{}
\author[5,6]{\fnm{M.J.} \sur{Way}}%\email{michael.way@physics.uu.se}
\author[7]{\fnm{Doris} \sur{Breuer}}%\email{}
%\author[9]{\fnm{Tim} \sur{Lichtenberg}}\email{}
%\author[10]{\fnm{Avi} \sur{Mandell}}\email{}
%\author[11]{\fnm{Jacob} \sur{Lustig-Yaeger}}\email{}
\author[8]{\fnm{Kanako} \sur{Seki}}%\email{}
\author[2,7]{\fnm{Caroline} \sur{Brachmann}}%\email{}
\author[9]{\fnm{Fabrice} \sur{Gaillard}}%\email{}
%\author[15]{\fnm{Renyu} \sur{Hu}}\email{}
\author[10]{\fnm{Manuel} \sur{Scherf}}%\email{}
\author[11]{\fnm{Svetlana V.} \sur{Berdyugina}}
\author[12]{\fnm{Brice-Olivier} \sur{Demory}}

%%\equalcont{These authors contributed equally to this work.}

\affil[1]{\orgdiv{Institute for Particle Physics and Astrophysics}, \orgname{ETH Zurich}, \orgaddress{\street{Wolfgang-Pauli-Strasse 27}, \postcode{8093} \city{Zurich}, \country{Switzerland}}}
\affil[2]{\orgdiv{Institute of Geological Sciences}, \orgname{Freie Universit\"at Berlin}, \orgaddress{\street{Malteserstr. 74-100}, \postcode{12249} \city{Berlin}, \country{Germany}}}
\affil[3]{\orgdiv{Division of Science}, \orgname{National Astronomical Observatory of Japan}, \orgaddress{\street{2-21-1 Osawa, Mitaka}, \postcode{181-8588} \city{Tokyo}, \country{Japan}}}
\affil[4]{\orgdiv{Earth-Life Science Institute}, \orgname{Institute of Science Tokyo}, \orgaddress{\street{2-12-1 Ookayama, Meguro-ku}, \postcode{152-8550} \city{Tokyo}, \country{Japan}}}
\affil[5]{\orgdiv{NASA Goddard Institute for Space Studies},  \orgaddress{\street{2880 Broadway}, \postcode{10025} \city{New York}, \country{USA}}}
\affil[6]{\orgdiv{Theoretical Astrophysics, Department of Physics and Astronomy}, \orgname{Uppsala University},\city{Uppsala}, \country{Sweden}}
\affil[7]{\orgdiv{DLR, Institute of Space Research},  \orgaddress{\street{Rutherfordstrasse 2}, \postcode{12489} \city{Berlin}, \country{Germany}}}
\affil[8]{\orgdiv{Department of Earth and Planetary Science} \orgname{University of Tokyo}, \orgaddress{\street{Komaba 4-6-1}, \city{Meguro-ku}, \postcode{153-8904}, \state{Tokyo}, \country{Japan}}}
\affil[9]{\orgdiv{Institut des Sciences de la Terre d'Orléans},\orgname{Université d’Orléans-CNRS-BRGM}, \orgaddress{\city{Orléans}, \country{France}}}
\affil[10]{\orgdiv{Space Research Institute}, \orgname{Austrian Academy of Sciences}, \orgaddress{\street{Schmiedlstrasse 6}, \city{Graz}, \postcode{8042}, \country{Austria}}}
\affil[11]{\orgdiv{Istituto ricerche solari Aldo e Cele Dacc\'o (IRSOL)}, \orgname{Faculty of Informatics, Universit\'a della Svizzera italiana}, \orgaddress{\street{Via Patocchi 57}, \city{Locarno}, \postcode{6605}, \country{Switzerland}}}
\affil[12]{\orgdiv{Center for Space and Habitability}, \orgname{University of Bern}, \orgaddress{\street{Gesellschaftsstrasse 6}, \city{Bern}, \postcode{3012}, \country{Switzerland}}}


\abstract{
The atmospheric composition of rocky exoplanets offers an important tool for constraining the properties of the interior of this type of planet, beyond what is possible from measurements of their mass and radius alone. 
However, the interpretation of these observations requires an understanding of the complex interplay of a larger number of coupled planetary and atmospheric processes. 
This review provides an overview of the current state of knowledge regarding rocky exoplanet atmospheres, beginning with their formation and escape mechanisms. We specifically highlight the importance of long-term interaction between the atmosphere, the surface, and the interior on rocky planets.
Furthermore, this review addresses the influence of biological activity and photochemical reactions on the atmospheric compositions.
Consequently, establishing how these different processes contribute to shaping the atmospheres of rocky exoplanets during their evolution is fundamental for the characterization of these planets with future space missions and ground-based surveys.
}


\maketitle
\section{Introduction}
Three decades after the discovery of the first exoplanet around a Sun-like star by \citet{1995MayorJupiter}, the focus of the field has shifted from the mere detection of exoplanets to their detailed characterization. 
The atmospheres of rocky exoplanets are of particular interest.

This review provides a summary of the major findings on the formation, evolution, and properties of rocky exoplanet atmospheres. 

% In the recent past, many studies have therefore focused on individual processes and stages of the atmospheric evolution, such as the formation of secondary atmosphere, feedback mechanism between the interior and atmosphere of rocky planets, or the role of the redox state of the planet. \todo{Should we add references here?} 
%This review 
\subsection{Atmospheres in the solar system}
\begin{figure}[ht]
    \centering
    \includegraphics[width=0.9\textwidth]{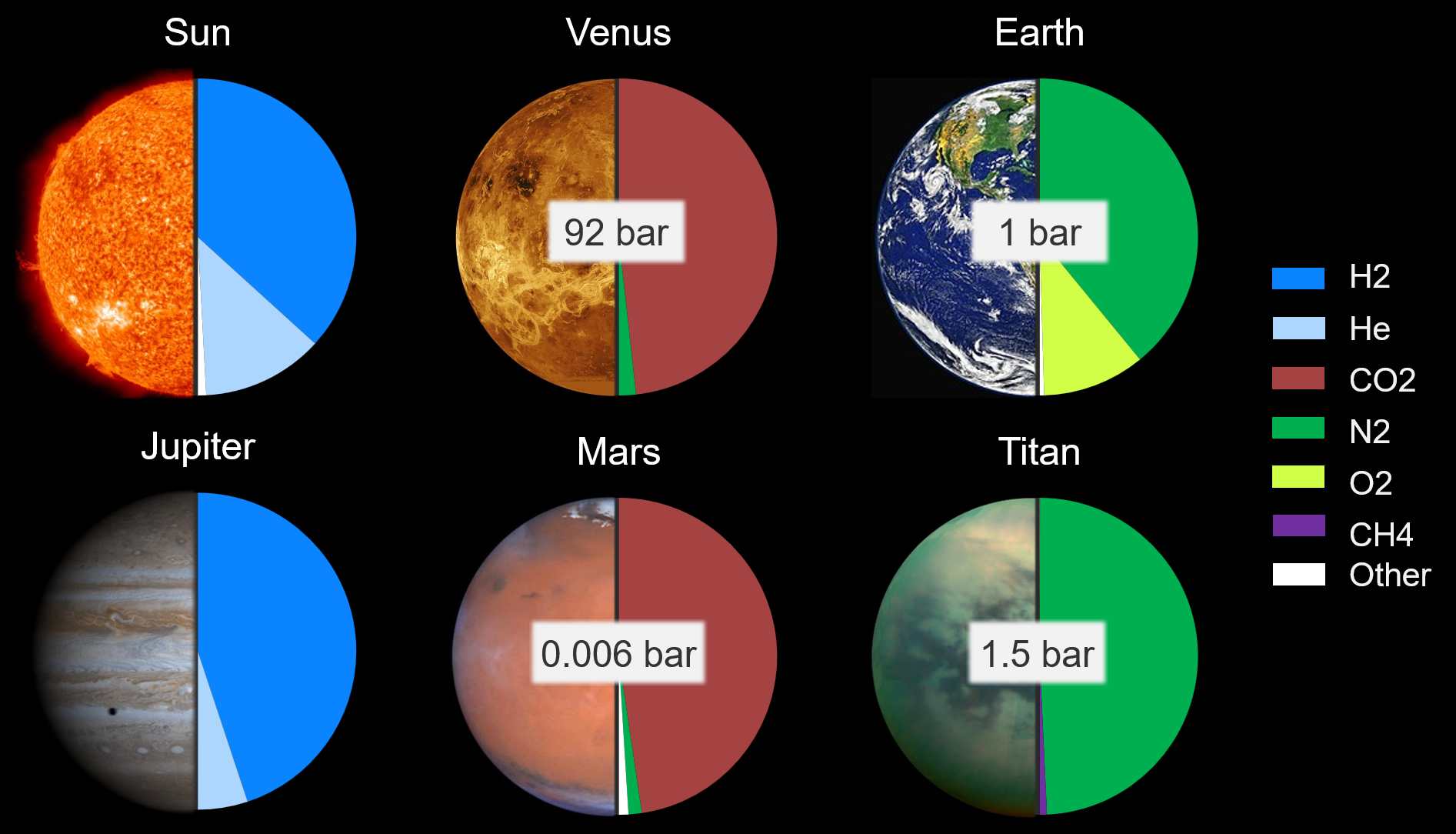}
    \caption{There are three main types of atmospheric compositions observed in the solar system: primordial atmospheres (Sun and gas giants), carbon-dioxide-dominated atmospheres (Venus and Mars), as well as nitrogen-dominated atmospheres (Earth and Titan). %\todo{Suggested for picture from Lena, of course we should make it more beautiful if we want to have something along these lines - any help is more than welcome!}
    }
    \label{fig:types_atms}
\end{figure}

% Lena, Marie-Luise
Our understanding of planetary atmospheres is mainly derived from laboratory experiments as well as from the observed diversity of atmospheres in the present-day solar system (Fig. \ref{fig:types_atms}). The composition of the atmospheres of the gas and ice giants (Jupiter, Saturn, Uranus, and Neptune) is, at least with respect to their main components, remarkably similar to the solar atmosphere. This indicates their primordial origin from the planetary nebula during the accretion phase of planets, where the main gases available in the system (hydrogen and helium, as well as trace amounts of other gases) were gravitationally captured by the accreting planets \citep[e.g.][]{hayashi1979earth}. 
For the rocky planets in the Solar System, it is generally believed that they either never accreted a primordial atmosphere or that they lost their primordial atmosphere during the early stages of their evolution via different mechanisms \citep[e.g.,][]{Lammer2018Atmospheres}.

Venus and Mars, the two neighboring planets of our Earth, both harbor a \ce{CO2}-dominated atmosphere, despite their different locations at the inner and outer edge of the habitable zone \citep[e.g.][]{kasting1993HabitableZones}, and their different planetary masses (Venus being almost as massive as Earth, but Mars only one tenth of the mass of Earth). The striking difference between these two atmospheres is the surface pressure -- with Venus' atmospheric pressure being about 100 times the pressure of Earth, and Mars' about 1/100. Several studies \citep[e.g.][]{Jakosky2018MarsLoss,2018Sakai_effects,Scherf2021,Lichtenegger2022,Hu2022Mars,thomas2023constraints} have tried to link the thin atmosphere on Mars with atmospheric loss processes and its magnetic field (see  Section \ref{sec:formationANDloss}), but it is still a topic of active debate.

The third type of atmosphere observed in the present-day solar system is nitrogen-dominated and appears on four very different bodies -- on Earth and Titan, and, as collisionless exospheres, on Pluto and Triton. Titan has a reduced chemistry nitrogen-dominated atmosphere with trace amounts of methane,  \citep[e.g.][]{mackenzie2021titan}, while on Earth, the nitrogen-dominated atmosphere together with a strong oxygen contribution is linked to the existence of life \citep{schirrmeister2015cyanobacteria,Spross2021} and the invention of photosynthesis. Pluto and Triton, in contrast, both have tenuous \ce{N2}-dominated exospheres with minor amounts of CO and \ce{CH4} that originate from vapor-pressure equilibria with their icy surfaces \citep[e.g.,][]{Scherf2025}. The question of how Earth's atmosphere would look like today if life had never evolved on Earth is a topic of active debate, especially in light of the discovery of rocky exoplanets in the habitable zone (HZ) around other stars. Although a lifeless Earth is likely similar to the late Hadean/early Archean eon, there are indications of similar or lower atmospheric pressure levels of nitrogen than today. For example, \citet{Marty2013} used isotopic data for 3.0-3.5 Ga to find that \ce{N2} was 0.5--1.1 bar. \citet{Avice2018} found at 3.3\,Ga that p\ce{N2} was similar or lower than today. Using a variety of different methods (e.g., fossil raindrops, lava vesicle sizes, oxidation of micrometeorites) at $\sim$2.6--2.7\,Ga, the total pressure could have been as low as 0.23 bar \citep{som2012air,Som2016,Rimmer2019Pressure}. Higher \ce{CO2} partial pressures in the Archean \citep{catling2020archean} are supported by the work of \citet{Kanzaki2015CO2,Charnay2017,Lehmer2020CO2,Feulner2023}. The evolution of the post-accretion period of Earth's atmosphere -- or of any of the rocky bodies' atmospheres -- remains poorly understood. Studies have produced a wide range of possible atmospheres (from highly reduced to oxidized) in the magma ocean phase \citep[e.g.][]{bower2019linking,2024Maurice} for which there are few constraints. For the other rocky bodies in the solar system, there is little information on the nature of the atmosphere during the accretion and magma ocean stage, although Titan's nitrogen isotope indicates that its atmosphere mainly comes from the accretion of \ce{NH3} ice and complex N-bearing organics \citep{Miller2019,Erkaev2021}. 

\begin{figure}[h!]
    \centering
    \includegraphics[width=\textwidth]{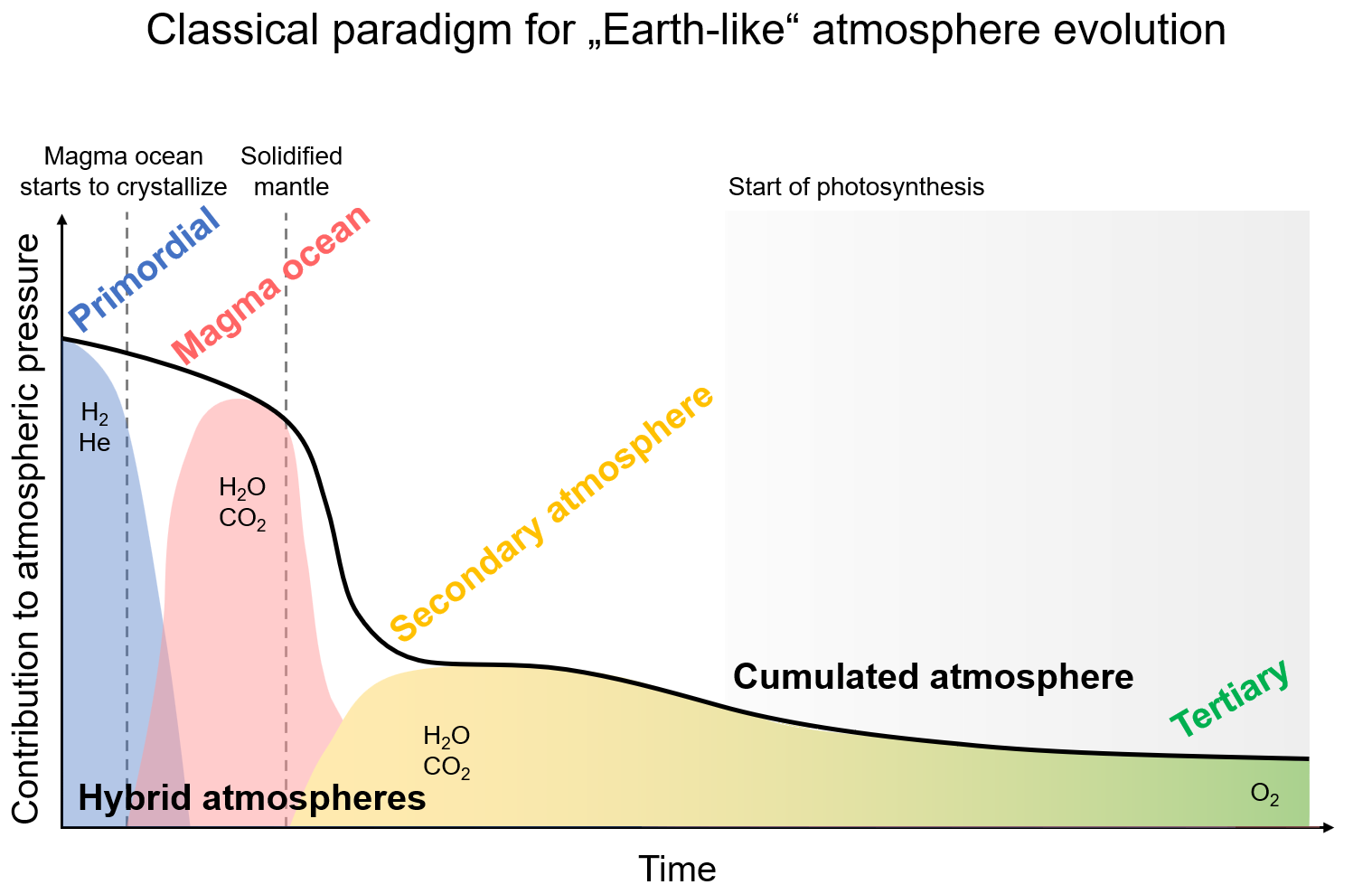}
    \caption{Sketch for the atmospheric evolution of an Earth-like planet from accretion (primordial atmosphere) via magma ocean crystallization and outgassing to a secondary atmosphere shaped by volcanic outgassing, or (in the case of Earth) by life (tertiary atmosphere). %\todo{From Lena, we should make it more beautiful - any help is more than welcome!}
    }
    \label{fig:atms_evol}
\end{figure}

Secondary outgassing by volcanic activity shapes the atmosphere of rocky planets during their long-term evolution. Note that for geological studies, the term secondary outgassing often does not encompass the magma ocean phase, whereas from an astronomer point of view as well as within this review chapter (see Fig. \ref{fig:atms_evol}) a secondary atmosphere includes any outgassing from the interior including dissolution of volatiles from the magma ocean \citep[e.g.][]{bower2019linking,2024Maurice}. It should be noted that the atmospheric composition can differ from the composition of volcanic gases expected to be degassed at the surface of a planet based on the local melt composition and redox state, as gases may remain dissolved in the melt based on their solubility depending on the already existing atmospheric partial pressures \citep[e.g.][]{gaillard2014theoretical,Gaillard2022,Sossi2023}. 

The atmosphere can change over geological timescales through different processes. These may include material in-fall (e.g., meteorites, comets), various atmospheric escape processes \citep{2020Gronoffescapeprocesses}, condensation of different species, weathering processes at the surface \citep{walker1981NegativeFeedback}, or even large volcanic outgassing \citep[e.g.][]{Wignall2011,Bond2017}. In the case of Earth, biology led to an oxygen-rich atmosphere after the Great Oxygenation Event more than 2.5\,Gyr ago \citep[e.g.][]{alcott2019stepwise}. We term this the tertiary atmosphere in Fig. \ref{fig:atms_evol}.

\subsection{Atmospheres of exoplanets}
\label{sec:atmos_exo}
% Lena, Marie-Luise
Beyond the solar system, different direct and indirect methods exist to detect and characterize planetary atmospheres. The first detection of an exoplanetary atmosphere was achieved for the hot Jupiter HD209458b using transit spectroscopy by \citet{2002_Charbonneau_sodium}, who measured an additional dimming of the star in the sodium resonance doublet by about 200 parts per million (ppm) as compared to the nearby continuum. This technique has become a leading approach for detecting exoplanetary atmospheres and characterizing their gaseous composition.

Whether an exoplanet is a gas giant or rocky planet can be determined indirectly based on their size, via transit spectroscopy, and the bulk density, via mass estimation using radial velocity measurements. Based on current observations, alongside theoretical work, rocky planets may be roughly limited to $<1.5\,R_\oplus$ \citep{2015RogersRocky,Fulton2017,2022Petigura_CKS}, although this value has been challenged by other statistical analyses of mass-radius data yielding values of 1.23$^{+0.44}_{-0.22}$ \citep[e.g.][]{chen2016probabilistic}. In general, the radius gap is mostly a population-level property, and classifying individual planets by their location relative to the gap is to be treated with caution \citepalias[see e.g, the discussion in][this Topical collection]{baumeister2025FundamentalsInterior}. 
The location of the radius gap further depends strongly on the stellar type and the orbital period \citep{2022Petigura_CKS,2023Horadiusvalley,2024_Parc_from,2024Venturini}.
Outliers have been repeatedly detected, and for each assumed transition mass/radius from rocky to sub-Neptune planets (with large fractions of volatiles in the form of gases, liquids and/or ices), there will be individual planets of smaller mass/radius with an extended radius (likely not rocky), such as L 98-58b \citep{2021Demangeon_Warm} or Kepler-138 d \citep{2023Piaulet_Evidence}, as well as larger planets with high densities indicating rocky planets. There is the suggested super-Mercury Kepler-107c with (based on current observations) $10\,M_\oplus$ and $1.6\,R_\oplus$ \citep{bonomo2023cold} leading to a density even higher than expected for a scaled-up Mercury planet. This suggests very small fractions of silicate materials mixed with heavy metals, while leaving almost no room for a substantial atmosphere. Rocky super-Earths and gaseous sub-Neptunes 
%(that have rocky cores embedded in a gaseous envelope with possibly liquid or solid layers underneath the gas envelope) 
may actually share a common origin, with different atmospheric loss processes dividing the sample into two subgroups (see \cref{sec:observationalfeatures}).

Another indirect indication of an atmosphere comes from the evaluation of flux variations during the orbit (phase-curves). In particular, measurements of the rise and ebb of an exoplanet's flux can provide constraints on a potential atmosphere. Depending on the wavelength, exoplanet phase-curves may probe predominantly thermal or reflective components of the planet's outgoing radiation, or their combination \citep[e.g.,][]{2021Wongexoplanet}, 

The first detection of thermal radiation from an exoplanet \citep{2005_Deming_thermal} and subsequent infrared phase-curve measurements \citep{2007_Knutson_phasecurve} allowed for the detection of clear atmosphere signatures on hot Jupiters. The thermal phase-curve amplitude was found to be significantly smaller than the planet's thermal emission measured during the occultation by the star (eclipse). This flux difference pointed to a significant part of the heat being recirculated from the day-side to the night-side of the planet. This recirculation was further corroborated by the East-ward shift of the phase-curve peak induced by the characteristic strong super-rotation occurring in the atmospheres of tidally-locked planets \citep{2002_ShowmanGuillot_atmcirculation}. Direct imaging studies of young gas giants also indicate cloudy and dusty atmospheres \citep[e.g.,][]{2010_Marois_directim}.

Optical phase-curves inform about the atmosphere's reflective properties, such as the wavelength-dependent geometrical albedo and the presence of hazes and clouds. For example, Rayleigh scattering on gas-phase atoms and molecules and small particles (haze, aerosols, etc.) produces a characteristic blueward slope in transit spectra. This was first found on the hot Jupiter HD189733b using transit spectroscopy at wavelengths longer than 550\,nm  \citep{2008_Pont_hd189733}. A high albedo of this exoplanet in the blue (at 450\,nm and shorter) was first determined using polarized-light optical phase-curves \citep{2008_Berdyugina_pol, 2011_Berdyugina_pol} and subsequently confirmed by near-UV secondary eclipse spectroscopy \citep{2013_Evans_hd189733}.
Thus, the Rayleigh slopes and high geometrical albedoes of exoplanets may indicate the presence of the atmospheres, hazes and clouds \citep[e.g.,][]{2013_Demory_clouds,2016_Sing_Rayleigh}.

The same general principles apply to the smaller rocky exoplanets. While there are hints of an atmosphere surrounding 55\,Cnc\,e \citep{2017_Mahapatra_55cnc_models,hu2024secondary}, the definitive evidence for an atmosphere around a rocky planet orbiting a cool star is still lacking. The absence of such atmospheres have been confirmed in several cases, either from thermal phase-curve measurements \citep[e.g.][]{2019_Kreidberg_LHS3844} or occultations for those rocky exoplanets subject to high instellation \citep[e.g.][]{2025_August_LHS1478, 2025_Meier-Valdes_TOI1468}. For more temperate planets (Venus-like or slightly hotter), such as TRAPPIST-1b and TRAPPIST-1c, current observations are still compatible with both an airless or atmosphere scenario \citep[e.g.][]{2023GreenTrappist1b,2023Ziebatrappist1c,Ducrot2025}.

The true characterization of the atmospheric chemical composition needs spectroscopic measurements to identify individual species in the atmosphere --  via transmission, reflection or emission spectroscopy. 
%Variations in the stellar spectrum during a planetary transit (primary eclipse) can be extracted from the baseline stellar spectrum, as photons passing through the transiting planet's upper atmosphere may show additional adsorption features that are not visible in the pure stellar spectrum (e.g., when the planet is behind the star from the observer point of view). Comparing stellar spectra (obtained during the secondary eclipse) with the light emitted from a planet (outside secondary eclipses) also reveals signatures of gases in exoplanetary atmospheres.
For gas giants and sub-Neptunes, these methods have been successfully applied to detect a variety of gases such as Na I, \ce{H2}O, \ce{CH4}, \ce{CO2}, CO, and \ce{SO2} \citep[e.g.,][]{2002_Charbonneau_sodium,
belu2011primary, 2018Wakeford_complete,benneke_jwst_2024, 2024_Powell_SO2, 2024_Beatyy_Sulfur,2024KemptonTransiting}.

Scattering on particles (aerosols) in exoplanetary atmospheres, such as cloud condensations and photochemical hazes, reduces the contrast of spectral signatures of gas-phase molecules, that makes them challenging to detect \citep[e.g.][]{fauchez2019impact}. In some extreme cases, even flat spectra have been observed, strongly suggesting the presence of aerosol-rich atmospheres 
\citep[e.g.,][]{2014_Kreidberg_superEarth_clouds, 2021_Gao_aerosols_review}. Overall, observations by {\em Hubble}, {\em Spitzer} and JWST reveal that aerosols are common in exoplanetary atmospheres, but disentangling particle sizes, composition and their height distribution in models remains a challenge due to multiple degeneracies.

%\todo{no such detection for rocky planets} 

Featureless or heavily muted transmission spectra of lower-mass exoplanets (sub-Neptunes/super-Earths) is a pattern that emerged in early observations \citep[e.g.,][]{2014_Kreidberg_superEarth_clouds} and consistently strengthened by the current JWST data and atmospheric retrievals \citep[e.g.,][]{fauchez2019impact, 2025_Madhusudhan_subNeptunes_JWST}. Along with the detection of multiple carbon-bearing molecules (\ce{CH4}, \ce{CO2}) and molecules previously hidden by aerosols, JWST data expose new inconsistencies that aerosol models must explain, in particular the complexity and heterogeneity of these atmospheres. For example, high metallicity, non-equilibrium chemistry \citep[e.g.,][]{2025_Jaziri_neq_chemistry} as well as organic and graphite hazes in carbon-rich atmospheres \citep[e.g.,][]{2025_Li_hazes_models} can explain some of the observed features.

In the future, spectra from rocky planets at large angular separations from their host stars can be directly observed in the optical and infrared using imaging coronography \citep[e.g.,][]{2024_Wolff_Roman_SPIE}, nulling interferometry \citep{quanz2022large}, etc. A detailed overview of different current and future observational capabilities from space and ground is given in \citet[][this Topical collection]{2026Lagage}.
%}
\section{Formation and loss processes of atmospheres}\label{sec:formationANDloss}
% \textbf{Caroline, Doris, Kanako, Marie-Luise, Tim, Philipp, Keiko, Renyu, Fabrice, Michael, Manuel, Kanako}
\subsection{Formation}
\label{sec:formation}
\Cref{fig:atms_evol} shows that the atmosphere of a rocky planet can be separated into primordial and secondary atmospheres. 
The primordial atmosphere (\cref{ssec:primordial}) is formed from gas of the accretion disk during planet formation. 
The secondary atmosphere (\cref{ssec:secondary}) is composed of volatiles that are initially brought to the planet with colliding planetesimals, impacting asteroids and comets, and are later released by various outgassing processes: During impact outgassing, which occurs particularly during accretion, volatiles are released directly into the atmosphere and do not accumulate in the forming planet at all. 
For other outgassing processes such as magma ocean outgassing and volcanic outgassing, the volatiles that are initially stored in the planet's interior during accretion are released later.

\subsubsection{Primordial atmospheres}
% Marie-Luise
\label{ssec:primordial}
Modern planet formation predicts that rocky planets form at least partly in the protoplanetary disk of the host star \citep[see e.g., reviews by][]{2022RaymondMorbidelli,2023Drkazkowskaplanetformation}.
These disks are mainly composed of gaseous hydrogen and helium \citep{2011Armitage}. 
If the thermal energy of a gas particle in the vicinity of a protoplanet is not enough to overcome the gravitational pull from the protoplanet, the gas particle is bound to the planet. 
The radius at which the thermal energy of a gas particle and the gravitational energy of the protoplanet are roughly equal is called the Bondi radius \citep{2006IkomaGenda}.
Thus, any planet for which the Bondi radius is larger than the radius of the planet itself binds gas from the surrounding protoplanetary disk \citep{2012IkomaHori,2014Lee,2016Ginzburg}. 
As the gas around the protoplanet cools and contracts, more and more gas will be gravitationally bound to the protoplanet \citep{2006IkomaGenda,2015LeeChiang,2016Ginzburg}. 
This minimum mass is small enough that rocky planets that form within the life-time of the protoplanetary disks will bind a primordial atmosphere. 

It is generally assumed that this primordial atmosphere is in hydrostatic equilibrium and consists of an inner convective and an outer radiative region \citep{2014PisoYoudin}. 
At the outer boundary, the Bondi radius, the pressure and temperature correspond to the local disk conditions. 

The opacity of the atmosphere plays an important role in shaping the profile of the atmosphere by determining the location of the radiative-convective boundary. There are two potential opacity sources, the gas itself and absorption by dust grains \citep{2003Semenovopacity}. The opacity contribution of dust grains depends on their size, as larger grains contribute less to the opacity. 
However, both analytical and numerical models of dust growth in primordial atmospheres, show that the contribution of dust grains to the opacity is low \citep{2014Mordasiniopacity,2014Ormelopacity}. Therefore, primordial atmospheres can be assumed to be grain-free, with the gas opacity being the dominant opacity source.

\begin{figure}
    \centering
    \includegraphics[width=\textwidth]{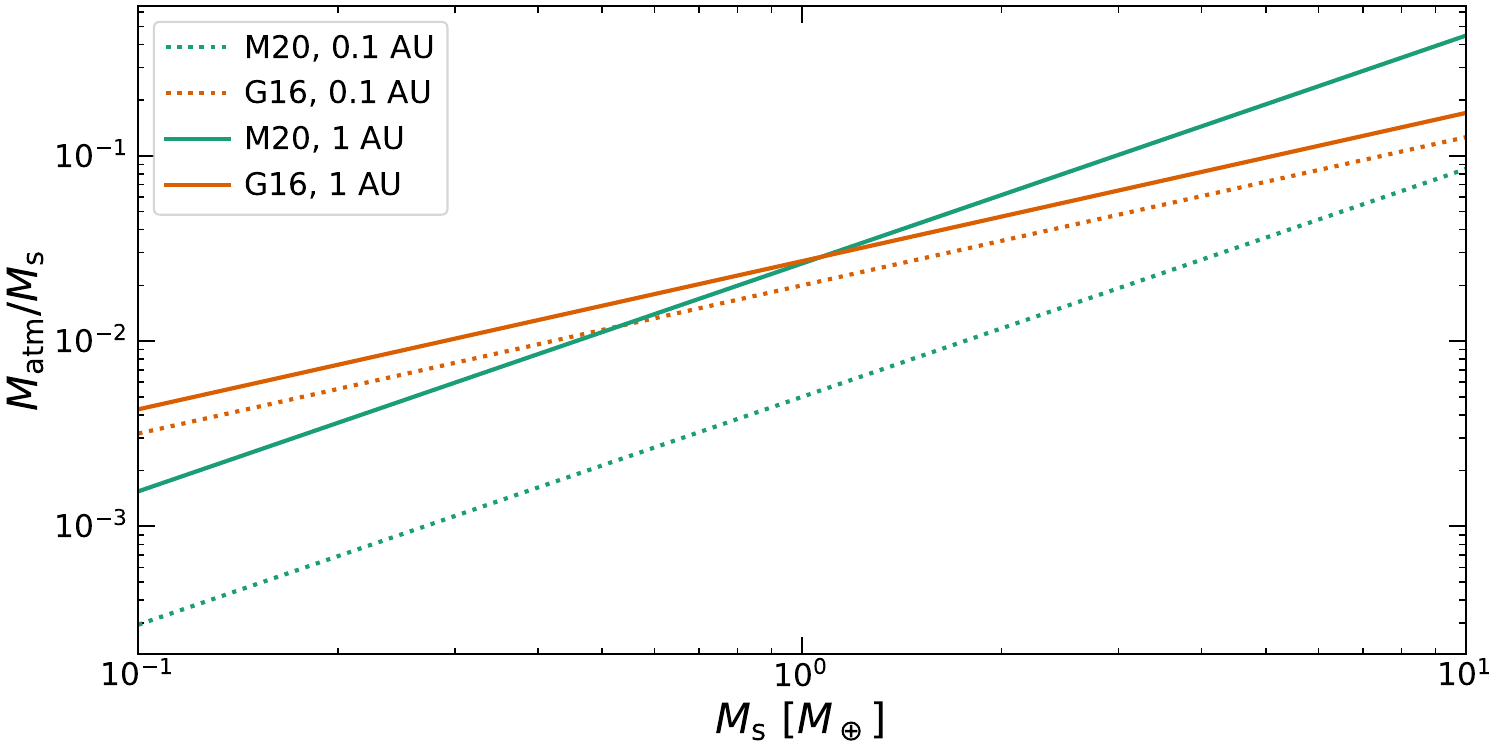}
    \caption{Comparison of the analytic scaling law for the primordial atmospheric mass fraction ($M_\mathrm{atm}/M_\mathrm{s}$) by \citet{2016Ginzburg} (orange line) to the fit to numerical simulations by \citet{2020Mordasini} (green line). Dotted lines refer to planets with an orbital distance of $a=0.1\,\si{\au}$ and solid lines to $a=1\,\si{\au}$. In all cases, the mass fraction of a primordial atmosphere of an Earth-sized planet is $\sim 1\,\%$, while a $10\, M_\oplus$ planet has a primordial atmospheric mass fraction of $\sim 10\,\%$.}
    \label{fig:primatm}
\end{figure}

The total accreted mass of a primordial atmosphere depends both on the local disk conditions and the optical properties of the atmosphere itself. \Cref{fig:primatm} compares an analytic estimation of the atmosphere mass fraction as a function of planet mass by \citet{2016Ginzburg} to the numerical fit by \citet{2020Mordasini}. \citet{2016Ginzburg} assume that the atmosphere mass is concentrated in the convective region and find 
\begin{equation}
    \frac{M_\mathrm{atm}}{M_s} \approx 0.02 \left(\frac{M_s}{M_\oplus}\right)^{0.8}\left(\frac{T_\mathrm{eq}}{10^3 \,\si{\K}}\right)^{-0.25}\left(\frac{t_\mathrm{disk}}{1\,\si{\Myr}}\right)^{0.5}.
\end{equation}
Here $M_\mathrm{atm}$ is the atmospheric mass, $M_\mathrm{s}$ the mass of the solid planet, $T_\mathrm{eq}$ the equilibrium temperature, and $t_\mathrm{disk}$ the life-time of the disk. 
\citet{2020Mordasini} on the other hand, use a population synthesis approach to find the mean primordial atmosphere mass fraction as a function of the solid planet mass and the orbital distance $a$
\begin{equation}
    \frac{M_\mathrm{atm}}{M_s} = 0.005 \left(\frac{M_s}{M_\oplus}\right)^{1.23}\left(\frac{a}{0.1\,\si{\au}}\right)^{0.72}.
\end{equation}
Both scaling relations estimate that the final primordial atmospheric mass fraction of an Earth-sized rocky planet is $\sim 1\, \%$ whereas a $10 M_\oplus$ super-Earth can have a primordial atmospheric mass fraction of $\sim 10\,\%$.

Initially, the composition of the primordial atmosphere corresponds to the composition of the protoplanetary disk. 
However, if the growth of the planet is dominated by smaller bodies (`pebbles'), the heating of the atmosphere by the release of gravitational energy leads to an enrichment of the atmosphere in heavier species due to the sublimation of the incoming bodies \citep{2014Lambrechts,2017Alibert,2018Brouwers,2023Steinmeyer}. 
Fragmentation and ablation of the accreted bodes can further enrich the atmosphere in elements other than H and He \citep{2011HoriIkoma,2016Venturini,2024MolLous}. 
The long-term evolution of such polluted primordial atmosphere is still uncertain \citep{2024Vazanrainout}.

One important consequence of atmospheric pollution is that it lowers the critical core mass to values as low as a few Earth masses \citep{2011HoriIkoma,2015VenturiniCoremass}. 
This is the critical core mass at which a planet enters run-away gas accretion and turns into a gas giant. 
The existence of super-Earths and sub-Neptunes demonstrates that it needs to be possible to avoid this runaway growth. 
Possible mechanisms for this are late formation \citep{2016LeeChiang}, interrupted accretion due to a gap opening further out in the disk \citep{2018FungLee,2019GinzburgChianggap}, and high opacity atmospheres \citep{2014Lee}.  For close-in planets, it has been proposed that the dominant mechanism to avoid run-away gas accretion is the recycling flows between the atmosphere and the surrounding protoplanetary disk \citep{2015OrmelRecycling,2022Moldenhauerrecycling}. 

While rocky planets will lose their primordial atmospheres during their later evolution, see \cref{ssec:escape}, these initial atmospheres can leave imprints on the planet's interior.
The combination of the heat from accretion and the heat produced by radioactive decay leads to the formation of a deep magma ocean. 
Hydrogen and noble gases such as He and Ne can readily dissolve into the magma ocean and later be sequestered in the mantle and core. 
The measured \ce{^{3}He}/\ce{^{4}He} ratio in Earth's deep mantle is considered evidence for the equilibration between a primordial atmosphere and this early magma ocean \citep{1996HarperJacobsen3He}.
The dissolution of hydrogen into the magma ocean also changes its redox state as discussed in \cref{ssec:secondary}.
Furthermore, the sequestration of hydrogen into the interior has been proposed as a mechanism to explain the observed bulk densities of super-Earths \citep{2022SchlichtingYoung,2024Rogers}. 
Beyond simple dissolution, chemical reactions between the primordial atmosphere and the magma ocean can further lead to an endogenic production of water with a D/H ratio of nebular gas \citep{2006IkomaGenda,2018Ikomawaterpart}. 

\subsubsection{Secondary atmosphere}
\label{ssec:secondary}
% Mike, Keiko Hanamo, Doris
\paragraph{Impact degassing}
\label{ssec:atmosphereimpact}
When a planetary body collides, a shock wave is generated at the point of impact and propagates through the impactor and the rocks near the impact site. The compression caused by the shock wave is irreversible, increasing entropy and converting part of the impactor's kinetic energy into heat. This shock heating induces phase transitions such as degassing, melting, and vaporization, and releases atmophile elements contained in solids (both the impactor and the target's surface) directly into the atmosphere as gases. This process, known as {\it impact degassing}, is considered to have played an important role in shaping planetary atmospheres throughout planetary formation, from the continuous accretion stage of planetary embryos \citep{1986MatsuiAbeimpactdegassing,1986Tyburczyimpactdegassing,2019Olsonnebular,2023Salvador} to the late accretion phase on fully formed planets \citep{1990Chybaimpactdelivery,2019Sakuraba_impact,2020Zahnleimpactatm}. The large number of craters preserved on the ancient surfaces of the Moon, Mercury and icy satellites attests the high frequency of collisions in the early solar system, suggesting that impact degassing played a significant role in the evolution of young terrestrial planets.

%\DB{comment: the next part is especially for water, keep it or write more general? ---- }
The efficiency of impact degassing to devolatilize an accreting planetary body and form an atmosphere increases with the amount of accretion heating, which in turn is related to the size of the growing body and the size of the impactor. As a planet grows, the impact velocities increase due to the increase in radius until partial and then, at larger radii, complete devolatilization of especially water occurs. Water in minerals is released due to the resulting high temperatures. At what planetary radius devolatilization becomes effective also depends on the hydrated minerals, and estimates for complete volatilization range from 1300 km to 2500 km \citep[e.g.][]{TYBURCZY2001}. However, experiments have shown that up to 30\% of volatiles can be stored in impact melts and projectile survivors \citep{Daly2018}. 
Impact outgassing during accretion is particularly important, as this process determines how much of the volatiles — especially water — is trapped in the forming planet. This can then alter the atmosphere in later stages of magma ocean and volcanic outgassing.

Once the volatile elements are released as gas, the gas composition in the hot vapor cloud in the initial phase is expected to be close to the chemical equilibrium composition due to the high temperature. Chemical equilibrium calculations with bulk elemental abundances similar to those of primitive meteorites indicate that the redox state of iron (iron oxides v.s. metallic iron/iron sulfides) and the amount of atmophile elements are particularly important for the composition and redox state of impact-generated gases \citep{Hashimoto2007,2007SchaeferFegleyoutgassing,2010SchaeferFegleychemistry}. The gas composition rapidly changes in accordance with the temperature and pressure changes associated with the vapor expansion, and eventually quenches at some point. The quenching conditions are determined by the cooling rate of an impact-induced vapor and the chemical reaction rates, and vary according to gas species. For instance, \citet{2020Zahnleimpactatm} used a thermochemical kinetics code to estimate that the quenching temperature of \ce{NH3} is about 300~K higher than that of \ce{CH4} for a Vesta-sized impact. In general, the quenching temperature tends to be lower as the size of an impact vapor cloud increases, due to a slower decrease in the internal temperature and pressure.

Ancient Mars is the poster child for an atmosphere by impact degassing. Presently, Mars has a thin atmosphere with global yearly mean surface temperatures of $\sim$--65\,C. However, there is ample geologic evidence that Mars in the Noachian eon ($>$ 3.7\,Ga) had warm periods (whose length is still debated) where substantial water flowed on the surface \citep[e.g.][and references therein]{Wordsworth2016}.
The seeds of the idea go back at least to \cite{Hashimoto2007}, who proposed that impact degassing could have contributed to Earth's atmosphere toward the end of its accretion period, leading to an atmosphere that may have been mainly composed of impact material. \cite{Haberle2019} proposed that a similar mechanism may have occurred on Mars, but in the post-accretion period. The theory is that the impactor creates a short-lived but hot thermal plume and that extant water on the surface of Mars interacts with the impactor's reducing materials to produce \ce{H2} (a powerful greenhouse gas). The authors estimated that if the impactor is $>$100\,km in size, it could provide enough \ce{H2} for the planet's mean global surface temperature to rise above 0\,°C if it already has a dense \ce{CO2} atmosphere. Earlier work by \cite{Wordsworth2017} had shown using ab-initio calculations that \ce{CO2}--\ce{H2} collisionally induced absorption (CIA) had previously been underestimated and could provide significant warming. Subsequent 3-D General Circulation Modeling (GCM) results confirmed that $\sim$10--20\% \ce{H2} in a 1--2 bar \ce{CO2}atmosphere could provide global and/or local surface temperatures above the freezing point in the Noachian and Hesperian eons \citep{Kamada2020,Guzewich2021,Schmidt2022}.
However, more recent empirical work by \cite{Turbet2020} has demonstrated that the \ce{CO2}--\ce{H2} CIA ab-initio calculations of \cite{Wordsworth2017} may be too optimistic and that the warming provided may be much less. Subsequent 3D GCM simulations by \cite{Schmidt2025} confirm this.

\paragraph{Magma ocean outgassing}
\label{ssec:magmaoutgassing}
% Fabrice, Mike, Lena
Magma ocean outgassing and solidification is an increasingly active field of research, as it rules the formation of both the initial secondary atmosphere and primordial reservoirs of volatile elements in the planetary mantle. Thus, it defines the starting point from which the secondary atmosphere and mantle reservoirs may evolve via enduring geodynamic processes.

Magma ocean outgassing is the process wherein the atmophile elements dissolved in magma oceans are exsolved to the surface, whereas when surficial gases are dissolved in the magma ocean, one talks about ingassing. In conditions where the exchange of materials between the atmosphere and the magma is most efficient, the mass and composition of the overlying atmosphere can be controlled by gas solubilities into silicate melts. An efficient exchange between the melt and the atmosphere occurs in particular during fractional crystallization. In this process, the heavier crystals sink and the lighter, more volatile melts rise and come into contact with the atmosphere through effective convection in the magma ocean, where the volatiles then exsolve or are ingassed \citep[e.g., ][]{nikolaou2019factors,2022Bowerwaterretention,2023Salvadorconvective}. In conditions of highly reducing surfaces, the atmosphere consists of \ce{H2} and \ce{CO}/\ce{CH4}, while N is dissolved in magma as nitrides, resulting in an atmosphere that is relatively N-depleted. Conversely, under oxidizing surface conditions, the predominant components are \ce{CO}/\ce{CO2} and \ce{N2}, while the majority of H is dissolved in magma as \ce{H2O}, resulting in an atmosphere that is relatively H-depleted. 

So far, the effect of magma ocean oxidation (or redox) state on outgassing has mostly been considered as a free parameter, which arbitrarily varies between end-member cases, that is to say, from very reduced to very oxidized \citep{2022Gaillardredox,2022Bowerwaterretention,2024Nichollsredox}. 
Oxygen fugacity f$_{\mathrm{O}_2}$ is a proxy that is used to quantify the redox state (~amount of oxygen) in the considered system and geochemists have used the iron-wustite (Fe-FeO, labelled IW) as a reference redox buffer. The most reduced systems have $\Delta$IW-6, which indicates oxygen fugacity values that are 6 orders of magnitude lower than the f$_{\mathrm{O}_2}$ imposed by the IW buffer. An example of such a reduced system is the mantle of present day Mercury. The most oxidized systems in contrast usually have $\Delta$IW+6, e.g., subduction related magma on Earth.  

The provenance of planetary accretion materials, both capture and escape of \ce{H2}, and internal redox processes contribute to controlling the magma ocean redox state.
The nature of the building blocks of planet accretion, pebble of chondritic origin, planetesimals and planetary embryos, is an obvious parameter controlling magma ocean f$_{\mathrm{O}_2}$. For example, in the solar system, a planet exclusively built from enstatite chondrites should be more reduced, whereas a planet composed of carbonaceous chondrites should be much more oxidized. Similar types of building blocks, with equivalent redox differences, may be expected in other planetary systems \citep{doyle2019oxygen}. Any intermediate magma ocean oxidation states can be attained by a mixture of these end-member materials. The disk processes controlling the provenance of these building blocks \citep{2015Morbidellisolarsystem,2021Johansenpebble} and what controls their redox state is much debated \citep{2020Righterredoxprocesses}. 
Other processes, such as the gravitational capture of nebular gases, a process that has long been discussed in the literature for its consequences in terms of volatile abundances \citep[see review by][]{2018Ikomawaterpart} and its potential effect on the magma ocean oxidation state, remain little investigated \citep[e.g.,][]{2023Young}. In all cases, the capture of the solar nebula would raise the \ce{H2}/\ce{H2O} ratio and thus drive the system toward more reducing conditions. It remains to be studied what type of mineralogy assemblies and exotic speciation could be generated in the planetary interior due to the accretion of H2-dominated primordial atmospheres as is the case for sub-Neptunes. Alternatively, atmospheric escape processes tend to favor H-loss, which should have the opposite effect, i.e., driving an increase in oxygen fugacity \citep{pierrehumbert2010palette, 2020Katyaloxidation}. The effect of such escape processes on the magma ocean oxidation state has also been little addressed in a quantitative way.

Internal oxygen redistribution, due to high pressures (> 15\,GPa) prevailing deep in the magma ocean, can also affect the oxidation state of the magma ocean. Most identified processes, that is to say, iron disproportionation \citep{2004Frostironmantle} and enhanced stability of ferric iron \citep{2024Zhangferriciron}, both favored by high-pressure, and silicon incorporation in the core, favored by high temperature, tend to cause an increase in the magma ocean oxygen fugacity. On the other hand, high temperature enhances the direct dissolution of oxygen in the metallic core, which causes a reduction of the oxygen fugacity of the magma ocean. 
All these processes are equilibrium processes. As such, they can be short-circuited in the chain of processes forming planets if non-equilibrium processes dominate.

Magma ocean solidification also strongly influences the initial distribution of water and other volatiles in the interior. As a certain amount of water is partitioned into the solidifying minerals, an exponential increase of water in the crystals toward the surface can be expected in particular for fractional crystallization: since volatiles prefer to enter/remain in the melt, the rising melt becomes increasingly enriched with volatile elements as the magma ocean solidifies over time. As a result, the volatile content also increases in materials that crystallize later.  The evolving magma ocean liquid is also continuously enriched in oxidized iron as Mg-rich silicates crystallize first. Therefore, Mg-rich cumulates can be found at the core-mantle boundary while Fe-rich cumulates form at the end of magma ocean solidification and remain close to the surface \citep{elkins2012magma}. This leads to a volatile and non-monotonic density increase toward the surface, resulting in a gravitationally unstable configuration, initiating early mantle convection and mixing the volatiles into the interior.

As described above, fractional crystallization suggests efficient degassing. A depletion of more than 90\% of the initial amount of volatiles has been assumed, and even small initial volatile contents (0.05\,wt.\% \ce{H2O}, 0.01\,wt.\% \ce{CO2}) can produce atmospheres in excess of 100 bars \citep{elkins2008linked}. However, fractional crystallization and degassing are disturbed among others by inefficient crystal-melt segregation, i.e., melt in which incompatible elements are concentrated can be confined in the solid cumulates. The faster the cooling and the slower the compaction, the less efficient is the crystal-melt separation \citep[e.g.,][]{2000Solomatovfluiddynamics,2017Hier-MajumdervolatileEarth}. The melt fractions that are trapped in the cumulate pile have been estimated to values of between 1 and 10\% \citep{elkins2008linked,2017Hier-MajumdervolatileEarth}. Furthermore, if the magma ocean did not comprise the entire mantle, the lower mantle remains primordial and thus more volatile-rich in comparison to the outgassed upper mantle due to magma ocean solidification \citep{Gaillard2022}. This means that a substantial amount of volatiles may still be present within planets after this early accretion and differentiation process. This water is then outgassed together with \ce{CO2} and other volatiles during subsequent volcanism, as is described in the next section.

\begin{figure}
    \centering
    \includegraphics[width=1\linewidth]{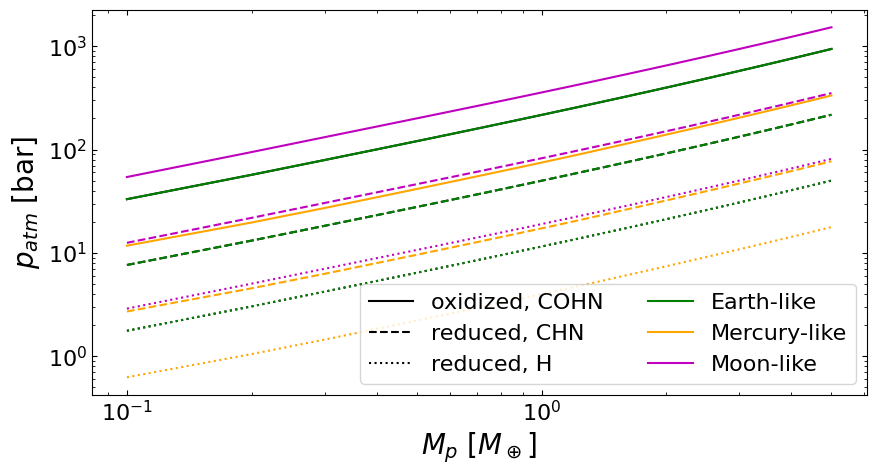}
    \caption{Rule-of-thumb estimates on maximal outgassed atmospheric pressure over planet mass considering different planetary interior configurations. The initial volatile concentrations in the mantle are 200 ppm \ce{CO2}, 200 ppm \ce{H2O} and 20 ppm \ce{N2}. For more information, we refer the reader to the main text. %\todo{Lena: this is what I had in mind, using example Earth-like/Moon-like/Mercury-like planets with the same initial mantle volatile concentration (here 200ppm CO2, 200ppm H2O and 20ppm N2), if we assume that all is outgassed (not taking into account solubility considerations), how would pressure change with mantle thickness and planet mass, compared then two a case where only CHN (CH4,H2,NH3) are outgassed (all O remains in interior to form silicates) or if C and N also are stored in mantle and only H2 is outgassed, how does atmospheric pressure change. Important: this is only to get better understanding of volatile budget, pressure are not realistic pressures due to solubility! Lena: y-axis Patm max, 10 Masses }
    }
    \label{fig:MO-atm-variation}
\end{figure}

The density of such an outgassed atmosphere (neglecting here condensation and atmospheric loss processes) can vary immensely depending on planetary mass, initial volatile content of the outgassing magma ocean, and magma ocean redox state. 
As an example, Fig.~\ref{fig:MO-atm-variation} shows the maximum outgassed atmospheric pressure versus planet mass for initial volatile concentrations of 200 ppm \ce{CO2}, 200 ppm \ce{H2O} and 20 ppm \ce{N2} in the mantle. We consider here three different planetary interior structures (with a small metal core comparable to the Moon, with an Earth-like interior structure, and with a Mercury-like extremely large metal core) assuming either an oxidized magma ocean (solid lines, outgassed species are \ce{CO2}, \ce{H2O} and \ce{N2}, i.e., leading to a COHN-rich atmosphere), a reduced magma ocean (dashed lines, outgassed species considered here are \ce{CH4}, \ce{H2} and \ce{NH3}) and an extremely reduced magma ocean where all C and N would remain stored in the mantle and only \ce{H2} would be degassed (dotted lines). The planet mass and interior structure (e.g., core-mass fraction $X_{CMF}$) influence most importantly the mass of the rocky mantle and therefore the total mass of COHN volatiles that can maximally be outgassed, $M_{\rm atm} \approx M_p(1-X_{\rm CMF})$, planet radius, 
\begin{equation*}
    R_{\rm p} = (7030 - 1840 X_{\rm CMF} \left(\frac{M_{\rm p}}{M_{\rm Earth}}^{0.282}\right),
\end{equation*}
and surface gravitational acceleration $g_0 \approx M / R_{\rm p}^2 \approx M_{\rm p}^{0.436}$ \citep{noack2020parameterisations}. 
% g = GM/R^2 ~ M/ [ M^(0.564) ] ~ M^0.436 

The resulting atmospheric pressure is then calculated via
\begin{equation*}
p_{\rm atm} [bar] = \frac{M_{\rm atm}g_0}{4 \pi R_{\rm p}^2} 10^{-5}  
\end{equation*}
% p ~ M^(1+0.436-0.564) = M^(0.872)
This simple rule-of-thumb calculation highlights that atmospheric pressures can vary by two orders of magnitude for the same planet mass but different interior configuration: 1) Mercury-like planets would tend to have a factor 3 smaller atmospheric pressure due to the reduced proportion of (volatile-bearing) rocky material; 2) The planet mass impacts the atmospheric pressure by a factor of almost 9 when increasing mass by a factor of 10 (not taking into account solubility effects); and 3) oxidized magma oceans lead to much denser atmospheres, with atmospheric pressures almost five times higher compared to the reduced case and more than ten times higher compared to the extreme case scenario of only \ce{H2} being outgassed.

%Lena: rule-of-thumbs MO calculations of which gases are when dominant, compare to existing similar studies on that
This simple estimate should of course be taken with caution and only be considered for illustrative purposes to point out that magma ocean atmospheres may vary strongly in density and composition from planet to planet. 
This example also considered extreme magma ocean redox state scenarios, and for many rocky planets, the atmospheric composition may lie in-between the oxidized and reduced scenarios. Furthermore, the redox state of the magma ocean may change over time due to interior differentiation processes (estimates for Earth suggest an increase of the mantle/magma ocean oxygen fugacity by several orders of magnitude during the accretion, core formation, and magma ocean crystallization, \citealp{wood2006accretion}). In addition, outgassing of reduced gases such as \ce{H2} may have further oxidized the magma ocean of early Earth, \citep{sharp2013hydrogen}.

\paragraph{Volcanic outgassing}
\label{ssec:secondaryatmospheres}
%Volcanic outgassing: Caroline
After the solidification of the magma ocean, volatile degassing continues through volcanic activity. Volatiles such as carbon (C), hydrogen (H), oxygen (O), sulfur (S), and nitrogen (N) are stored within the rocks of the mantle and crust. Various geological processes, including heat transfer, decompression, and the introduction of water, can induce the melting of these rocks, leading to the release and partitioning of the stored volatiles into the melt. Being less dense than the surrounding rocks, the melt ascends through the mantle and crust, forming magma diapirs or utilizing fractures and cracks in the brittle crust to rise. As the magma ascends, the decreasing pressure reduces the solubility of dissolved volatiles and eventually leads to the formation of a gas phase when the magma becomes supersaturated. These gas bubbles are subsequently released into the atmosphere either through extrusive volcanism or, in cases where the melt is trapped, through cracks in the overlying rocks, driven by buoyant forces. The composition of these gas bubbles varies depending on factors such as the volatile content of the melt, oxygen fugacity, temperature, and degassing pressure. Melts with high redox states predominantly degas \ce{H2O}, \ce{CO2}, \ce{N2}, and \ce{SO2}, whereas at lower redox states, volatiles such as \ce{H2}, \ce{CO}, \ce{CH4} (especially at high pressures), \ce{S2}, \ce{H2S} and \ce{NH3} are more common \citep{scaillet2009sulfur, gaillard2014theoretical}. 
%\lena{
At low degassing (subaerial) pressures, sulfur is outgassed as \ce{SO2}, whereas at higher (submarine or subsurface) degassing pressures, sulfur is outgassed as \ce{H2S}, leading to different atmospheric compositions for planetary surfaces covered by oceans compared to atmospheres formed predominantly by continental volcanism \citep{gaillard2011atmospheric}.
%} 

Additionally, degassing of halogens like HCl and HF is possible and while highly toxic \citep{Giggenbach1996}, they tend to play a less significant role on a global scale.
Several studies have aimed to estimate variations in atmospheric composition, pressure, and temperature on exoplanets whose atmospheres are primarily formed through volcanic degassing. To accurately assess outgassing efficiency, key parameters such as the mantle's volatile content, melting fluxes, and oxygen fugacity must be taken into account. \citet{gaillard2014theoretical} highlighted the critical role of pressure in controlling the speciation of volcanic gases: within the pressure range of 50 to 500 bar, carbon species like \ce{CO2} and \ce{CO} dominate, while at pressures between 50 and 10\textsuperscript{-3}\,bar, \ce{H2O} becomes the most abundant species. At even lower pressures, sulfur-bearing gases prevail. Building on this, \citet{ortenzi2020mantle} investigated the influence of mantle redox state and planetary size on atmospheric pressure. Their results show that planets with reduced interiors generate thinner atmospheres compared to those with oxidized interiors, due to the higher molecular weight of oxidized volatile species. Moreover, they found that volcanic degassing is most efficient on planets with masses between 2 and 4 Earth masses. This is attributed to greater volatile inventories, increased radiogenic heat production, and elevated mantle temperatures. However, for planets exceeding this threshold, higher lithospheric pressures raise melting temperatures, thereby inhibiting or even preventing volcanic outgassing.

\citet{woitke2021coexistence} developed a simplified equilibrium chemistry model applicable to surface temperatures below 600\,K, predicting three distinct compositional atmospheric types for cold terrestrial planets. These types are defined by the relative abundances of four dominant gases: Type A atmospheres consist of \ce{NH3}, \ce{CH4}, \ce{H2O}, and either \ce{H2} or \ce{N2}; Type B feature \ce{O2}, \ce{H2O}, \ce{N2}, and \ce{CO2}; and Type C are characterized by \ce{H2O}, \ce{CO2}, \ce{CH4}, and \ce{N2}. Building on this framework, \citet{brachmann2025distinct} extended the model by incorporating mantle volatile partitioning and volcanic degassing. Their study revealed a strong dependence of atmospheric chemistry and pressure on mantle oxygen fugacity, surface temperature, and melt production rates. Mantles with oxygen fugacities below $\Delta$IW~+1 tend to produce thin, nitrogen-rich Type A atmospheres, while those with higher oxygen fugacities generate \ce{H2O}- or \ce{CO2}-dominated Type C atmospheres. Surface temperature further modulates gas abundances: \ce{H2O} and \ce{N2} are more prevalent at higher temperatures, whereas \ce{NH3} and \ce{CH4} dominate under cooler conditions. Notably, Type B atmospheres cannot be generated by volcanic degassing alone, as \ce{O2} is neither directly released nor formed through equilibrium chemistry. Instead, abiotic \ce{O2} formation requires photodissociation of \ce{H2O} molecules. 

In a complementary study focused on hotter planets with surface temperatures above 800\,K, \citet{liggins2022growth} introduced sulfur into the C-H-O-N system and identified three atmospheric classes governed by mantle redox state. Class R atmospheres (oxygen fugacity < $\Delta$IW~+0.5) are dominated by \ce{H2}, \ce{CH4}, and \ce{NH3}; Class I atmospheres (oxygen fugacity between $\Delta$IW~+0.5 and +2.7) include \ce{CH4}, \ce{CO2}, \ce{CO}, and \ce{COS}; while Class O atmospheres (oxygen fugacity > $\Delta$IW~+2.7) are rich in \ce{SO2}, \ce{S2O}, and polymeric sulfur species (S\textsubscript{x}). %\textbf{\todo{from Lena: this is a bit too much detail, as exact numbers also vary on input parameters. Can you generalize these different atmosphere types a bit more?}}

% \begin{table}
%     \caption{Composition of outgassed atmospheres \lena{I would leave this table out, as there are different studies with different classifications}}
%     \begin{tabular}{lcc}
%      Class    & Composition & Requirement \\
%          &  &
%     \end{tabular}
% \end{table}
It should be noted that the studies referenced here assume a homogeneous and temporally stable oxygen fugacity across a planet’s mantle, melt, and atmosphere. However, this is a rough approximation. In reality, oxygen fugacity can vary between these different reservoirs and is influenced by a variety of processes. For example, the degassing of \ce{H2} and CO may oxidize the melt, while sulfur degassing may reduce it \citep{burgisser2007redox, gaillard2014theoretical}. Additionally, atmospheric processes such as the photodissociation of \ce{H2O} and the subsequent escape of \ce{H2} can oxidize the atmosphere \citep{sharp2013hydrogen}.

The suggested atmosphere classifications therefore also depend on additional factors or processes not yet included, such as planetary composition, evolution of surface temperature under different weathering/condensation scenarios, and atmospheric losses depending among others on the host star. 

%\ML{volcanic outgassing in ocean on land differnt due to different pressures}

\subsubsection{Current limitations in our understanding of planetary atmosphere formation}
% Fabrice
Our understanding of planetary atmosphere formation is still constrained by significant approximations and assumptions that require rigorous reassessment. Several major limitations---both experimental and theoretical---persist, demanding focused attention to refine current models.

The degassing and ingassing processes within magma oceans are currently described by solubility laws extrapolated to extreme temperatures and redox conditions, yet these extrapolations lack robust experimental or thermodynamic validation. While widely used by the scientific community, such extrapolations introduce uncertainties that remain difficult to quantify. Moreover, the very nature of species dissolving in magma oceans is not fully understood, particularly for compounds such as \ce{H2}, \ce{NH3}, and \ce{CH4}.

Beyond internal chemistry, the coupling between the primordial and secondary atmospheres presents a modeling challenge. Recent models propose the emergence of hybrid atmospheres, formed by planetary outgassing into a primordial \ce{H2}-He envelope \citep{tian_atmospheric_2024}. The transition from primordial to secondary atmospheres has only recently been approached \citep{2024KrissansenTotton}. The subsequent ingassing of this primordial atmosphere into the planetary interior — and its geochemical consequences — remains to be constrained. A more versatile modeling approach, capable of integrating exotic conditions beyond standard paradigms, is essential to capture these complex dynamics.

The formation of the metallic core plays a critical role in shaping atmospheric composition through equilibration between the magma ocean and the metallic core. Many key atmophile elements — such as carbon, nitrogen, and hydrogen —also exhibit siderophile tendencies, complicating their partitioning between the core and the atmosphere \citep{2021Gaillardinoutgassing}. This duality necessitates finer modeling of distribution mechanisms during the evolution of the planet.

While the assumption of equilibrium during ingassing and outgassing is convenient, it has been increasingly challenged \citep{2025Walbec}. Non-equilibrium processes — such as bubble nucleation, volatile diffusion from melt to bubbles, and the balance between advection and convection in extreme magmatic conditions—require systematic study. Validating predictive equations under these high-pressure, high-temperature conditions remains a major experimental and theoretical challenge.

Planetary impacts represent another source of disequilibrium. They play a dual role in the evolution of the atmosphere by acting as volatile reservoirs and triggers for large-scale degassing, while at the same time potentially leading to partial or complete atmospheric loss. Despite their recognized importance, these stochastic events are not yet fully integrated into global models of atmospheric formation.

Finally, the solidification of magma oceans and associated degassing form a complex dynamic system, where internal convection and external radiation interact. Degassing influences the solidification process, which in turn affects convection rates and cooling rates, while solidification itself impacts degassing, creating feedback loops that remain poorly characterized. Additionally, the interplay between stellar irradiation and the nature of the outgassed atmosphere may lead to divergent evolutionary pathways \citep{hamano2013emergence}. Thus, magma ocean solidification, degassing, atmosphere formation, and stellar irradiation are deeply intermingled processes whose interdependencies require further elucidation.

\subsection{Escape}
\label{ssec:escape}
In general, thermal atmospheric escape occurs when the kinetic energy of gas particles in the upper atmosphere exceeds the gravitational binding energy of the planet. 
The upper atmosphere typically refers to either the region above the homopause or above the mesopause. 
In the heterosphere, above the homopause, the different atmospheric species are no longer mixed, leading to a stratification by mass. 
The region above the mesopause is characterized by a positive thermal gradient with altitude. 
The two boundaries are typically located close to each other. 
The height of the mesopause on both Earth and Venus is at $\approx 100\,$km, for exoplanets, the boundary between inner and upper atmosphere is $5-25\,$\% bigger than the transit radius \citep{2020KiteBarnettenvevolution,2020MalskyRogersevolutionenvelopes}.

Several physical processes can contribute to increasing the energy of the atmospheric particles. 
This section serves as a brief summary of important mass loss mechanisms. 
For a detailed discussion on atmospheric escape, we refer the reader to \citet[][this Topical collection]{chapter8}. 

\subsubsection{Thermal escape processes}
\label{ssec:thermalescape}
% Contributors: Manuel, Marie-Luise
In the case of thermal mass loss, the heating of the planetary atmosphere by external or internal sources drives the escape of the atmosphere. 
% \todo{add Jeans parameter}.
% \ML{something about composition effects}
We distinguish two different regimes of thermal escape. In Jeans escape, the upper atmosphere is in hydrostatic equilibrium, and the fraction of the escaping particles corresponds to the high-energy tail of a Maxwell distribution \citep{1955jeans,1963Chamberlein}.
For strong heating, the upper atmosphere is no longer in hydrostatic equilibrium.
The  resulting expansion of the atmosphere leads to a bulk outflow of atmospheric gas similar to solar Parker winds \citep{1965Parker}. In contrast to Jeans escape, the outflowing gas remains a collisional fluid. 
This regime is referred to as hydrodynamic escape.
The transition between Jeans and hydrodynamic escape typically takes place when the kinetic energy is the same or greater than the gravitational binding energy of the planet.

The heating source plays an important role in shaping the mass loss rate and the final structure of the planet. In the following, we discuss the two common heating sources for thermal escape. 
\paragraph{Boil-off and core-powered mass loss}

Both in the so-called boil-off and core-powered mass loss, the main drivers of the mass loss are the thermal energy reservoir of the planet and the stellar bolometric irradiation\footnote{Meaning the irradiation across all wavelengths-.} \citep[e.g.,][]{2016Ginzburg,2016OwenWu,2024Tangcore}. 
The stellar irradiation heats the atmosphere from above at the same time as the heat flux from the solid planet heats the atmosphere from below. 

While the planet is embedded in the disk, the pressure support from the surrounding  protoplanetary disk and the solid accretion luminosity allow the primordial atmosphere of the planet to be extended out to its Bondi radius \citep[e.g.,][]{1996Pollack,2000Ikomaformation,2014PisoYoudin,2015Stoekl_hydrodynamic}. 
\citet{2016OwenWu} find that the atmospheres of low-mass planets remain inflated after the disk disperses, as the cooling time of the atmosphere is longer than the timescale of the disk dispersal. 
The resulting loss of hydrostatic equilibrium establishes a bulk outflow of atmospheric gas in the form of a hydrodynamic wind \citep{2016OwenWu,2016Ginzburg}. 

Boil-off ends when the mass loss timescale becomes shorter than the cooling timescale of the planet \citep{2024Rogersboiloff,2024Tangcore}.
The typical timescale of boil-off is on the order of a few Myrs, making it a very fast loss process \citep{2016OwenWu,2024Tangcore,2024Rogersboiloff}.
The remaining mass fraction depends on the solid planet mass and the strength of the irradiation \citep{2016OwenWu,2024Tangcore}. 
Models show that boil-off can strip the total primordial atmosphere of planets with $M_\mathrm{p} \lessapprox 5\, M_\oplus$ that receive a flux of $F\gtrsim100\,F_\oplus$ \citep{2017Fossati,2024Tangcore}. 

The combination of the thermal energy of the solid planet and the bolometric irradiation from the host star continues to drive atmospheric escape over a timescale of Gyrs. This type of atmospheric escape is known as core-powered mass loss.  
However, most works on the role of core-powered mass loss on the evolution of planets have used a simplified analytical model based on \citet{2018Ginzburg}. 
Recent numerical models by \citet{2024Tangcore} find that the analytical model overestimates the mass loss due to core-powered mass loss. 
Further work by \citet{2025Misenernonisothermal} finds that the mass loss rate depends significantly on the ratio of the opacity in the visible and infrared wavelength range. 
Therefore, boil-off and core-powered mass loss likely only play a role for planets with primordial atmospheres. 

\paragraph{Photoevaporation}
%Leads: Marie-Luise
Photoevaporation refers to the hydrodynamic escape of the upper atmosphere driven by the EUV and X-ray (XUV) irradiation received from the host star \citep[e.g.,][]{2012Lopez,2014Lammer,2012OwenJackson,2013OwenWu}. 
The XUV irradiation heats the upper atmosphere via dissociation and ionization of gas molecules.
As the XUV luminosity is especially high for young stars \citep[][this Topical collection]{2005Ribas,Johnstone2021Stars,chapter8}, photoevaporation plays a significant role in the loss of the primordial and magma ocean outgassed atmosphere in the first $\sim 100\,$Myr \citep[e.g.,][]{2009Gilmannvenusearly,2017OwenWuradiusvalley,Lammer2018Atmospheres}.

Photoevaporation can lead to a complete loss of the primordial atmosphere of rocky planets if the initial mass-loss timescale $t_\mathrm{loss} = M_\mathrm{atm}/\dot{M}_\mathrm{atm}$ is shorter than the phase of high XUV output from the star \citep{2017OwenWuradiusvalley}. 
For close-in planets at roughly $0.1\,$AU, this is the case for planets up to $6\, M_\oplus$ \citep{2013OwenWu,2020Mordasini}.
The duration of the phase with high XUV output, also referred to as the saturation phase of the star, decreases with increasing stellar mass \citep{2012Jackson,2014ShkolnikBarmanMdwarfXUV,2019McDonaldXUVstars,Johnstone2021Stars}.
In addition, for a given equilibrium temperature planets orbiting young M dwarfs receive a higher XUV flux compared to planets orbiting FGK stars as they are on shorter orbits \citep{Johnstone2021Stars}.
Consequently, photoevaporation is more efficient around low-mass stars. 
Rocky planets around M dwarfs are therefore less likely to retain atmospheres than planets around more massive stars. 

Modeling the variety of physical and chemical processes influencing the escape rates due to photoevaporation is computationally expansive, which is why the escape rates are often estimated using the energy-limited approximation \citep{1981Watson,2007Erkaev}, i.e.,
\begin{equation}\label{eq:en-lim}
     \dot{M}_\mathrm{atm} = \eta  \frac{\pi R_\mathrm{XUV}^{2}R_\mathrm{pl}}{4 \pi G M_\mathrm{pl} K},
\end{equation}
where $G$ is the gravitational constant, $\dot{M}_\mathrm{atm}$ is the atmospheric mass loss rate, $F_\mathrm{XUV}$ is the stellar XUV surface flux at the planet, $R_\mathrm{XUV}$ is the effective radius where $F_\mathrm{XUV}$ is absorbed in the atmosphere, $R_\mathrm{pl}$ and $M_\mathrm{pl}$ are the planetary radius and mass, respectively, and $K$ accounts for Roche-lobe effects. The heating efficiency, $\eta$, is the fraction of the absorbed XUV surface flux that goes into heating. Besides atmospheric type (see below) and local conditions, it depends on the spectral energy distribution (SED) of the XUV flux, whereby a harder SED leads to an increased value for $\eta$ \citep{Linsky2025}. This implies that active young K and G stars, rapidly rotating stars, and active M dwarfs induce higher $\eta$-values.

In addition, $\eta$ depends on a planet's gravitational potential. Exoplanets with higher gravities have smaller atmospheric scale heights, implying an absorption of the XUV flux at lower atmospheric densities and a smaller value for $\eta$ \citep{Salz2016,Linsky2025}. In total, $\eta$ may vary between roughly 1\% and 20\% for hydrogen-rich atmospheres \citep[e.g.,][]{2014Shematovichefficiency,Salz2016}. For secondary atmospheres, however, heating becomes more complex and depends on various additional factors such as the net energy expanse of the various photochemical reactions and radiative cooling of additional cooling agents such as \ce{CO2} or CO in \ce{CO2}-dominated atmospheres \citep[e.g.,][]{2018Johnstone_upper,Yoshida2022,2025VanLooveren_Hazard} and atomic line cooling via oxygen in \ce{N2}-\ce{O2}-dominated atmospheres \citep{Nakayama2022}.

In the energy-limited approach, the mass loss rate is determined by the ratio of the incoming irradiation and the gravitational binding energy of the planet. 
However, it has been shown that the energy-limited formula overestimates the escape rate for close-in hydrogen-dominated, highly-irradiated low-mass planets, whereas it underestimates the escape rate for cool planets with more compact atmospheres \citep{2016Erkaevmassloss,2016OwenMohanty}. 
One solution to overcome this is to use pre-calculated grids based on hydrodynamical models \citep[e.g., ][]{2018Kubyshkinaescapegrid}. 

\subsubsection{Non-thermal mass loss}
% Contributors: Kanako Seki
In the upper atmosphere of terrestrial exoplanets, collisions become less frequent, and the suprathermal component in the velocity distribution can contribute to atmospheric escape. Two major drivers of non-thermal atmospheric escape are the stellar XUV irradiation and stellar wind. Non-thermal escape mechanisms are categorized into collisional non-thermal escape and stellar wind-induced escape. Photochemical escape and charge exchange (energetic neutral atom production exceeding escape velocity) are major collisional non-thermal escape mechanisms. Stellar wind-induced escape includes ion pickup, atmospheric sputtering, and cold ionospheric outflows (plasma instabilities) for unmagnetized planets, as well as polar wind, auroral outflows, and plasmaspheric drainage plumes for magnetized planets \citep[see Table 2 in][this Topical collection]{chapter8}. The relative importance of these mechanisms depends on planetary conditions, such as atmospheric composition and intrinsic magnetic field.

On Earth-like or heavier planets, where gravitational escape velocity is considerable ($11.2\,$km/s for Earth), thermal escape primarily concerns hydrogen. Heavier species like oxygen, carbon, and nitrogen need to be accelerated to reach escape velocities via non-thermal escape processes. Atmospheric escape of these heavier species, which often constitute a major part of the secondary planetary atmosphere of rocky planets, mainly occurs as ion escape. Neutral atoms or molecules in the upper atmosphere can be ionized by stellar XUV irradiation, charge exchange interactions, or electron impact. Various processes supply planetary ions to the magnetospheres \citep[e.g.,][and references therein]{2015Sekiplasma}.

For magnetized planets like Earth, ion escape primarily occurs from the polar ionosphere, corresponding to latitudes higher than the subauroral regions \citep[e.g.,][]{1993Abe_exos,2001Sekioxygenloss,2007JYau_polar}. Planetary ions outflowing from the polar ionosphere undergo various acceleration and transport in the magnetosphere, with a significant part eventually escaping to the interplanetary space \citep[e.g.,][]{2020Gronoffescapeprocesses}. Some ions, despite having energies above the escape energy, return to the planetary atmosphere due to magnetospheric convection and plasma processes such as pitch angle scattering by wave-particle interactions. Other escape mechanisms include plasmaspheric drainage plume and ENA production by charge exchange \citep[e.g.,][and references therein]{2006Keikachargeexchangeloss,2020Gronoffescapeprocesses}.

For unmagnetized planets, the stellar wind can interact directly with the planetary upper atmosphere. In addition to Jeans escape (thermal escape), photochemical escape, which we will briefly address in the context of an upper atmosphere's photochemistry in Section~\ref{sec:photo}, can be an important mechanism for neutral atmospheric escape \citep[e.g.,][]{Amerstorfer2017,2017JLillisMaven}. The relative importance of neutral escape compared to ion escape depends on stellar XUV irradiation and planetary mass. As planetary mass increases, the relative importance of ion escape to neutral escape generally becomes higher due to the large escape energy. Major ion escape mechanisms from unmagnetized planets include ion pickup including polar plumes \citep{2015Curyyionpickup,2022Sakakuraionpolarplume}, atmospheric sputtering \citep{1991Luhmannoxygenpickup}, and cold ionospheric outflows \citep{2019Inuiionoutflowsmars}. The cold ionospheric outflow involves various plasma processes causing ion acceleration/heating. Detailed description about the non-thermal escape processes and related observables can be found in \citet[][this Topical collection]{chapter8}.

\subsubsection{Stripping by giant impacts}
% Keiko Hamano
Giant impacts induce a strong shock wave that propagates through the impactor and the target planet, thereby inducing a global ground motion over them. The atmosphere overlying the surface gains momentum from the ground motion, and, if the propagating shock wave is sufficiently strong, it escapes from the planet on a global scale \citep[e.g.][]{1997ChenAhrens}. One-dimensional hydrodynamic simulations indicate that the loss efficiency by this mechanical process is strongly dependent on pre-impact surface conditions, specifically in the presence or absence of an ocean \citep{2005GendaAbe, 2015Schlichting,2024LockStewart}. Owing to the different shock impedances between the rock, water and gas, the velocity of the ocean surface becomes higher than that of the rock surface without an ocean. As a result, the presence of an ocean enhances the loss efficiency of the planetary atmosphere significantly. For example, on an Earth-mass planet with a 1 bar \ce{H2} atmosphere, in the absence of an ocean, the ground velocity would need to be greater than the escape velocity to completely remove the atmosphere. However, if there is a 3 km deep ocean, the atmosphere could be completely stripped away with a ground velocity of about 40\% of the escape velocity \citep{2024LockStewart}. The effect of oceans in enhancing atmospheric loss becomes stronger as the ocean-to-atmosphere mass ratio increases. In comparison to the surface conditions, the effects of other parameters such as atmospheric composition and surface temperature are relatively minor.
Quantifying the efficiency of atmospheric loss caused by giant impacts has been attempted by means of 3D SPH (smoothed particle hydrodynamics) simulations \citep[e.g.][]{2020Kegerreis}. However, this remains challenging due to difficulties in resolving the crust, atmosphere and ocean accurately.

In the aftermath of giant impacts, the planetary surface becomes extremely hot through the propagation of the strong shock wave and the re-accretion of the ejected material. Consequently, a high-temperature (exceeding several thousand Kelvins) mixed atmosphere would form on the post-impact surface, consisting of the remaining volatile elements and the partially vaporized silicates. The extremely hot atmosphere may expand against the planet's gravity and escape hydrodynamically into space. \citet{2021BierstekerSchlichting} considered that thermal radiation from the lower atmosphere drives hydrodynamic escape and estimated the associated energy-limited mass loss rate. They found that planets with masses that are less than half of the Earth's mass are capable of losing their H-He dominated atmospheres, as long as the mean molar weight remains sufficiently low, even in the presence of vaporized silicate. As other heavier gas species increase, the mass loss rate decreases rapidly with increasing mean molecular weight. A more precise estimation of the mass loss rate would require hydrodynamic calculations that account for the detailed structure and energy budget of escaping atmospheres.

\subsection{Observational features of escape}
\label{sec:observationalfeatures}
\subsubsection{Present-day Solar System}\label{sec:solarsystem}
% Mike, Manuel
While there is evidence that the terrestrial planets have undergone significant thermal mass loss in the past (see Section~\ref{ssec:escapeeffect}), it only plays a minor role at the present day and is limited to the lightest element, hydrogen \citep{Lammer2006,Jakosky2018MarsLoss}.
Instead, the dominating escape processes for the terrestrial planets in the Solar System at present-day are non-thermal processes, such as the escape of ionized species and photochemical escape. Below we will summarize some key observations of these escape mechanisms; for more details, see, e.g., \citet[][this topical collection]{2015Chappellionsphere,Ramstad2018,Scherf2021,Gillmann2022,chapter8}.

Atmospheric escape at Venus is dominated by ion escape for oxygen and carbon, and by photochemical escape for hydrogen. Ions with a sufficient energy to gain escape velocity are produced either in Venus' exosphere and ionosphere via the Sun's XUV irradiation and are then accelerated by the solar wind or ionospheric electric fields \citep[e.g.,][]{Hartle1990,Luhmann2004,2017Slapak_atmospheric}. Escaping ions were, for instance, observed with the ion mass analyzer (IMA) of the ASPERA-4 instrument onboard Venus Express (VEX), see, e.g., \citet{2007Barabashescaperate}, \cite{Fedorov2011}, and \cite{2018PerssonVenusescape}. Observations between 2006 and 2014 in the induced magnetotail of Venus yielded escape rates for \ce{O+} and \ce{H+} of $\sim 2.9\times10^{24}$\,s$^{-1}$ and $\sim 7.6\times10^{24}$\,s$^{-1}$, respectively, for solar minimum and somewhat lower values for solar maximum due to an increase of the proton return flux toward the planet during solar maximum \citep[see,][for details]{2018PerssonVenusescape}. However, the ion mass analyzer onboard VEX was only able to distinguish \ce{H+}, \ce{He+}, and heavy ions (e.g., \ce{C+} and \ce{O+}) from each other, implying that the derived \ce{O+} escape rate prescribes the total ion escape of all heavy ions added together \citep{2018PerssonVenusescape}. Recently, however, measurements with the Mass Spectrum Analyzer onboard Bepi-Colombo while flying by Venus indicate the \ce{C+} to \ce{O+} ratio to be up to $\sim$0.3 \citep{Hadid2024}.

Whereas ion loss is the only significant source of escape into space for O, photochemical escape provides the dominant sink for hydrogen on Venus \citep[][see also Section~\ref{sec:photo} for details]{Lammer2006,Chaffin2024}. Energetic photochemical reactions for both species, however, produce hot coronae, i.e., exospheric suprathermal particles around Venus, that can be observed in Ly-$\alpha$. Its hydrogen corona was already observed by Mariner~5 \citep{Anderson1976} and later by the Venera missions \citep{Bertaux1978,Bertaux1982}, the Pioneer Venus Orbiter (PVO) \citep{Cravens1980} and VEX \citep{Chaufray2012}. Its structure can be approximated via a two component model consistent of a cold, thermal and a hot, non-thermal component \citep[e.g.,][]{Chaufray2012} from which thermal and photochemical escape rates can be estimated, which are of the order of $10^{19}$ and $4\times10^{25}$\,s$^{-1}$ \citep[e.g.,][]{Lammer2006}, respectively. Recently, \citet{Weichbold2025} analyzed ion cyclotron wave observations by VEX in Venus' exosphere and were able to derive photochemical escape rates for H and, for the first time, D. For hydrogen, their derived value of $\sim4\times10^{25}$\,s$^{-1}$ is in agreement with Monte Carlo simulations \citep{Lammer2006,Chaffin2024}  and it highlights that photochemical escape is indeed the major source of H escape on Venus with loss rates being almost an order of magnitude higher than non-thermal H$^+$ escape \citep{2018PerssonVenusescape}. Oxygen, on the other hand, whose hot corona was for the first time observed with the UV spectrometer onboard PVO \citep{Nagy1981}, is too heavy to escape photochemically (see Section~\ref{sec:photo}). The energy that oxygen obtains from the dissociative recombination of \ce{O2+} \citep[i.e., $\leq$6.96\,eV; e.g.,][]{Kim1998} is smaller than the relatively high escape energy at Venus' exobase (i.e., $\sim$8.6\,eV as calculated through the gravitational potential); see also Section~\ref{sec:photo}.

On Earth -- the only terrestrial planet in the solar system that hosts both a collision-dominated atmosphere and an intrinsic magnetic field -- atmospheric escape is dominated via ion outflow along the open field lines of its polar ovals. Observations of the so-called polar wind, whereby atmospheric ions are accelerated upwards from the ionosphere along the magnetic field lines due to the charge separation electric field present in the ionosphere \citep{2015Chappellionsphere}, date back to the late 1960s and early 1970s when \ce{H+}, \ce{He+}, and \ce{O+} were for the first time observed to flow upwards from the ionosphere into the magnetosphere \citep{Shelley1972,Sharp1977,2015Chappellionsphere}. Various thermospheric, exospheric and magnetospheric missions collected data of the Earth's plasma environment since then  \citep[see, e.g.,][for an overview]{2020Dandouras}, and it was, for example, found through high-altitude spacecraft observations that only around 10\% of the upflowing \ce{O+} is indeed escaping via open field lines with the rest flowing back toward the Earth; a complex process that leads to a total upper limit for the \ce{O+} escape rate of $\sim5\times10^{25}$  \citep{2001Sekioxygenloss}. More recent observational studies found a positive correlation of the \ce{O+} loss on the solar wind dynamic pressure \citep{Haaland2012,Schillings2019}, as well as a significant increase of the escape during geomagnetic storms \citep{Haaland2012}; extreme solar conditions therefore seem to be able to push \ce{O+} outflow rates beyond $10^{26}$ \citep[e.g.,][]{Haaland2012,Wei2012CIR,Schillings2019}. \ce{N+} escape rates, on the other hand, are typically observed to be an order of magnitude lower than the \ce{O+} escape rates \citep{Lin2020}.

Other escape channels at the Earth are typically negligible, except H escape, which predominantly stems from photochemical reactions in the thermosphere. As is the case for Venus and Mars, Earth's hydrogen corona can be mapped via Ly-$\alpha$ observations, e.g., with the Solar Wind Anisotropies instrument onboard the Solar and Heliospheric Observatory (SWAN/SOHO) from which the loss rate of H can be estimated. This was found to be in the order of $10^{26}$\,s$^{-1}$, which -- if it were constant through geological history -- would remove around 1 meter of the terrestrial ocean into space over billion-year timescales \citep{Baliukin2019}. However, H escape toward the past has most likely been increasingly higher, specifically during the Archean and Hadean eons \citep[e.g.,][]{Lammer2018Atmospheres}. Another method to map Earth's extended H corona is to observe energetic neutral atoms (ENAs), e.g., with the Interstellar Boundary Explorer (IBEX) spacecraft, which are produced through charge exchange between exospheric hydrogen and shocked solar wind protons, predominantly at the subsolar magnetopause \citep{2010Fuselier_energetic}. This charge exchange process further generates X-rays planned to be observed by the Soft X-ray Imager (SXI) onboard the SMILE mission \citep{Sembay2025}.   

For Mars -- the smallest of the three planets with a collision-dominated atmosphere -- atmospheric escape is more diverse. Whereas photochemical reactions are the dominant loss channels for nitrogen, oxygen, and likely carbon, it is thermal escape for H and sputtering for heavier species. For oxygen, photochemistry is primarily driven by the dissociative recombination of \ce{O2+} and \ce{CO2+} \citep[][see also Section~\ref{sec:photo}]{2020Gronoffescapeprocesses}. Theoretical models, however, predict that not all the oxygen atoms will gain sufficient energy to escape but will form a corona of hot oxygen surrounding the planet \citep{WALLIS1978escape,Amerstorfer2017}. This corona was observed by the Mars Atmosphere and Volatile Evolution mission by measuring the O I 130.4\,nm triplet feature \citep{2015DeighanMaven, 2024ChirakkilMarsMaven} with escape rates being derived to be in the order of $10^{25}$ \citep{2015DeighanMaven}, i.e., comparable to O escape rates on Venus and Earth. 

For N, the dominant production channels are the photodissociation of \ce{N2} and the dissociative recombination of \ce{N2+} \citep[e.g.,][]{Fox1980,Cui2019}. Based on simultaneously modeling photochemical reactions and MAVEN observations of \ce{N2} and \ce{N2+} profiles, the photochemical escape rate of nitrogen was derived to be slightly below $10^{25}$\,s$^{-1}$ \citep{Cui2019}. Also for H, photochemical reactions provide an important escape channel, but, based on Ly-$\alpha$ observations of the Martian H corona with HST, this is only as high as roughly 30\% of the thermal H escape rate of $\sim10^{26}$\,s$^{-1}$ \citep{Bhattacharyya2023}. Here, the radiative dissociation of \ce{HCO+} provides the major photochemical loss reaction \citep{Gregory2023HCO}.

Photochemical escape further played an important role during the Martian history. The radiative dissociation of \ce{N2+} is likely responsible for the strong atmospheric fractionation of \ce{^14N}/\ce{^15N} \citep{Fox1997}, C, and even O \citep[e.g.,][]{Lillis2017,Amerstorfer2017}. Photochemical escape could have led to an integrated loss of \ce{CO2} from Mars during the last 4\,Gyr of about 200\,mbar \citep{Amerstorfer2017}, making it, together with ion escape, the most important \ce{CO2} loss mechanism \citep{Lichtenegger2022}. For the first 500\,Myr of the Martian evolution, however, thermal escape was the dominant driver of atmospheric escape \citep{Scherf2021}.

Sputtering is another relevant loss process from the Martian atmosphere that was until recently only observed indirectly, e.g., via measuring the precipitation of planetary heavy pickup ions onto the atmosphere, which subsequently sputter other particles into space \citep[e.g.,][]{2015LeblancMarsionprecip,2018Leblancsputtering}. Recently, however, first direct observations of sputtered argon at Mars were successfully attained with MAVEN by correlating Ar densities with the solar wind motional electric field, thereby deriving Ar sputter rates to be around $2\times 10^{23}$\,s$^{-1}$, a value that is in agreement with the higher end of modeled sputtering rates \citep{Lichtenegger2022}. Although sputtering only contributes relatively minor loss rates compared to other species and escape channels, it is the dominant loss mechanism for species heavier than O and can provide important insights for reconstructing atmospheric losses on Mars over time since it preferentially removes lighter isotopes from the atmosphere, i.e., \ce{^36Ar} compared to \ce{^38Ar} in this specific case \citep{Jakosky2018MarsLoss,Lichtenegger2022}. A final important loss channel on Mars is heavy ion escape (i.e., C, N and O ions), which was measured by Mars Express \citep[e.g.,][]{Lundin2013} and MAVEN \citep{Brain2015}, and can exceed $\sim10^{25}$\,s$^{-1}$ during high solar activity \citep{Ledvina2017}. Observations have further shown that the plasma structure around Mars is dominated by tailward outflows of ions \citep{2012NilssonMarsion,2019Inuiionoutflowsmars} and a permanent northern polar plume made of energetic ions that contributes more than 20\% to the total heavy ion escape \citep{2015DongMaven}.

Besides the aforementioned planets, one may also mention Titan with its dense \ce{N2}-dominated atmosphere. The Ion and Neutral Mass Spectrometer onboard Cassini, for instance, observed a suprathermal corona around the satellite, through which suprathermal escape rates for N and \ce{CH4} were estimated to be, again, in the order of $10^{25}$\,s$^{-1}$ \citep{DeLaHaye2007}. Sputtering, specifically induced by energetic particles in Saturn's magnetosphere, is another significant escape channel with loss rates in the order of $10^{25}$\,s$^{-1}$, whereas ion escape and particularly thermal escape are comparatively insignificant, mostly due to the low XUV surface flux received from the distant Sun \citep[see, e.g.,][for an overview on atmospheric escape at Titan]{Erkaev2021}.

As one can see from this summary, the loss rates from Venus, Earth, Mars, and even Titan are roughly on the same order of magnitude. This seems to be relatively surprising as it was initially thought that a planetary magnetic field would protect the atmosphere from escape by interaction with the stellar wind. Its protective role, therefore, remains highly debated \citep{Ramstad2018,Ramstad2021,Way2023}. While the diversion of energy from the stellar wind energy by a magnetic field protects the atmosphere from sputtering and ion pickup, the magnetosphere increases the interaction region between stellar wind and the planet, thus leading to a higher amount of transferred energy \citep{2018Gunellmagneticfield}. In addition, one can see that photochemical escape plays a crucial role for various species on all of these planets, a process whose efficiency is not necessarily diminished by an intrinsic magnetic field. Whether a magnetic field can provide shielding for present-day conditions and, even more so, for more intense solar/stellar plasma and radiation environments remains unknown.

\subsubsection{Exoplanets}
% Marie-Luise
In the case of exoplanets, the outflow of gas from the atmosphere can in principle be seen through wavelength-dependent signatures in the transit spectrum of a planet \citep[e.g., ][]{2023DosSantos}. 
The first such observation was the detection of escaping H from the hot Jupiter HD 209458 b \citep{2003Vidal-MadjaH209,2008EhrenreichHD209}.
%The escape of neutral hydrogen seen by a Lyman-$\alpha$ (Ly$\alpha$) feature. 
Since then, outflow of H has been detected in a number of planets including sub-Neptune-sized planets such as HD 63433 \citep{2022ZhangHD63433, 2023DosSantos}.

The escape of heavier elements such as carbon, nitrogen, and oxygen can be probed in the UV wavelength regime. 
However, these metal lines are typically very weak and have only been observed in hot Jupiters so far \citep[e.g., ][]{2010Linskyescape,2010Schlawinescape,2023DosSantos}. 
The UV spectroscopy capability of the upcoming HWO, however, will enable the study of atmospheric escape across a wide range of planets including cool sub-Neptunes \citep{2025DosSantosEscapeHWO}.

A promising observational probe of atmospheric escape is the helium triplet located at $1.083\,\mu$m. 
This line is observable both in transmission spectroscopy and at high spectral resolution with ground-based facilities \citep[e.g.,][]{2018Allartspectrally,2020Vissapragadaconstrains,2022Kirkkeck,2023Zhanggiant}.

We can further study atmospheric escape by examining its impact on the exoplanet population as a whole. 
One of the most significant results of the Kepler mission is the detection of the so-called radius valley or radius gap, which refers to a lack of planets with radii between 1.5 to 2.0$\,R_\oplus$ \citep[e.g., ][]{2017FultonKepler,2018VanEylenradiusvalley,2018FultonPetiguraradiusgap,2020Bergerradiusgap,2023Horadiusvalley}.

This bimodal distribution in planet size was actually predicted by theoretical models of photoevaporation previous to its detection \citep{2013OwenWu}. 
According to photoevaporation models, planets below and above the gap are part of the same population of planets with Earth-like interiors that are initially surrounded by a large primordial atmosphere. 
After formation, these planets undergo significant thermal mass loss.
Depending on their total mass, planets will either lose their entire atmosphere turning into super-Earths located below the radius gap or retain a fraction of the initial atmosphere, becoming sub-Neptunes located above the gap. 
Numerous studies have demonstrated that atmospheric escape processes described in \cref{ssec:thermalescape} can explain the observed features of the radius gap, such as its dependence on the orbital period and stellar mass \citep[e.g.,][]{2017OwenWuradiusvalley,2018JinMordasinievaporation,2019Guptacorepowered,2021Rogersphotoevapvscorepowered}. 

However, the radius gap can also be explained by two distinct planet compositions. 
In this picture, planets below the radius gap have a composition similar to Earth, while planets above the radius gap can contain up to 50 wt.\% water \citep[e.g., ][]{2020Venturinirvformation,2022LuquePallee,2024Burnwaterworld}. 
Although most models assume that super-Earths have undergone atmospheric escape, it is also possible that these planets formed without ever accreting a primordial atmosphere in the first place \citep{2021LeeConnersprimordialradiusgap,2022Leeradiusgap,202Nielsenprimordialrv}.

One strategy for testing the two competing scenarios, atmospheric escape and water-rich formation, observationally involves multi-planet systems that straddle the radius gap.
If the mass and radius are known for both planets, one can then test if the system architecture is consistent with predictions from atmospheric escape models or not \citep[e.g.,][]{2020OwenTesting,2024EggerUnveiling,2025LockleyTOI1117}
This will require a large sample of small-sized planets with precise mass and radius measurements \citep{2024LacadelliCharacterisation}. 
The upcoming PLATO mission is expected to discover a vast number of multi-planet systems \citep{rauer2025PLATOMission}.
Thanks to the planned radial velocity follow-up observations to obtain mass measurements, the PLATO sample will thus allow a more detailed characterization of the radius valley. 
Planets situated directly within the radius valley such as TOI-544\,b \citep{2024OsbornTOI544} are of particular interest.
Confirmation that the radius gap is caused by atmospheric escape implies that rocky planets typically accrete a primordial atmosphere. 
The potential influence of such an atmosphere on the long-term evolution of rocky planets, e.g., through interactions with the magma ocean, thus needs to be studied.
% \todo{
% - will the mass loss remian after plato}

\subsection{Effect of atmospheric loss on the evolution of rocky planets}
\label{ssec:escapeeffect}
% Marie-Luise, Lena
In the most extreme case, atmospheric escape can lead to the total erosion of the atmosphere, leaving behind a bare rock. 
This might be the case for close-in rocky planets around M-dwarfs \citep[e.g.][]{2025VanLooveren_Hazard}. 
Even in a less extreme scenario, atmospheric escape may drastically transform the composition of an accreted or outgassed atmosphere. 
\citet{luger2015extreme} suggested that it is possible to transform a sub-Neptune type planet into a rocky planet residing in the habitable zone. \citet{tian2014high,luger2015extreme,Meadows2017} suggested that planets orbiting early phase M-dwarfs or planets in a runaway greenhouse state could become strongly oxygen-rich over time, due to photodissociation and subsequent loss of hydrogen. 
\citet{carone2025co2} extended these studies by including \ce{CO2} in the magma ocean atmosphere, and found that the build-up of free \ce{O2} in the atmosphere is, to some extent, possible by atmospheric erosion. 
\citet{2025Cherubimoxidationgradient} show that the dissolution of water in the molten mantle shields the oxygen from escape (by limiting the amount of water in the atmosphere). Later water mantle outgassing can lead to oxidized secondary atmospheres \cite[e.g.][]{luger2015extreme,Meadows2017}.
This suggested oxidation of the atmosphere by erosion of reducing gases (especially of hydrogen) also fits the observations from the solar system (see also further down), where Mars shows a thin but \ce{CO2}-dominated atmosphere, whereas theoretical predictions would suggest a more reduced outgassed atmosphere \citep{scaillet2009sulfur}.  
% From Mike:
%({\bf I don't see how. The previous studies imply there would be a lot of O2 around, but in fact we see very little on Mars. Maybe this needs to be explained a bit more?})

For a high enough escape rate, hydrogen can drag along heavier elements \citep{1986ZahnleKasting,1987Hunten,2018OdertNoblegases}. 
This can lead to a loss of moderately volatile rock-forming elements on Mars-sized planetary embryos, altering their Fe/Mg and Si/Fe ratios compared to their host star \citep{2020Benediktrockvolescape}.
More importantly, atmospheric escape can modify the abundance of the radioactive isotope \ce{40K}, which is an important constraint on the tectonic evolution of a rocky planet \citep{2020Benediktrockvolescape,2023Erkaevradioactive}. 

For small, rocky planets on ultra-short orbits, the temperature on the dayside surface is hot enough to vaporize the planet \citep[e.g.,][]{kite2016atmosphere,2022ZilinskasObservability,2024CurryEvolution}. 
The evaporated rocky material then escapes into space through thermally driven winds \citep{2013PerezBeckerCatastrophic}. 
The mass loss rates are high enough to result in disintegration lifetimes down to a few hundred Myr \citep{2013PerezBeckerCatastrophic}.
To this date, four such planets have been found observed through asymmetric transit profiles and time-variable transit depths caused by comet-like tails formed by large dust grains: Kepler-1520\,b \citep{2012RappaportKIC}, KOI-2700\,b \citep{201Rappaport2700b}, K2-22\,b \citep{2015SanchisOjedaK2}, and BD+05 4868\,Ab \citep{2025Hondisintegrating}.
These planets offer a unique opportunity to directly probe the interior of rocky planets \citep{Curry2025}.

The strength of a past atmospheric loss can also be estimated from the fractionation of different isotopic pairs in the remaining atmosphere, most prominently for the deuterium to hydrogen ratio. In the case of Venus and Mars, it implies high losses of hydrogen (be it from \ce{H2} or \ce{CH4} in the atmosphere, \ce{H2O} steam, or liquid, evaporated water oceans). 
Other isotopic fractionation patterns can be observed as well, for example for xenon, carbon, nitrogen and argon \citep{zahnle2019strange,gillmann2011volatiles,2020Lammermodvolatilefrac}. 
The \ce{^36Ar}/\ce{^38Ar} and \ce{^20Ne}/\ce{^22Ne} isotope ratios of the Earth and Venus are comparable to solar ratios \citep{2000PorcelliPepinearlyearth,2001PorcelliEarthraregas,2004YokochiMartyNeon,2009Gilmannvenusearly,mukhopadhyay2012noblegases,2018OdertNoblegases,2019WilliamsMukhopadhyayNeons}. 
These isotope ratios are seen as evidence that proto-Venus and proto-Earth have captured a \ce{H2}-dominated atmosphere \citep[e.g.,][]{1980MizunoNeon}. 
In order to reproduce the measured noble gas ratios, \citet{2021LammerVenusEarthMarsformation} find that proto-Earth must have grown to a mass of $0.5-0.6\,M_\oplus$ during the disk's lifetime, while Venus accreted most of its mass before the disk disperses. 
However, these isotope ratios can also be reproduced through implantation of solar wind onto accreted material \citep{2003PodosekNoble,2016Moreiraorigin,2016PeronNeon}.
Nevertheless, the evolution of atmospheric composition is an important constraint on the formation timescale of terrestrial planets in the Solar System. 

\section{Interaction and feedback between atmosphere, surface, and interior}
% Phillip
Even though the atmosphere and mantle of a planet are not in direct contact anymore after the solidification of the primordial magma ocean, there is still interaction between the atmosphere and mantle. This is primarily facilitated by the transfer of volatile species between the different planetary reservoirs (mantle, crust, ocean, atmosphere). For example, volcanic outgassing from the mantle brings various gases into the atmosphere, while simultaneously depleting the mantle. Plate tectonics and crustal delamination can bring surface-bound volatiles back into the mantle (see \ref{ssec:secondaryatmospheres}).

\begin{figure}
    \centering
    \includegraphics[width=0.7\linewidth]{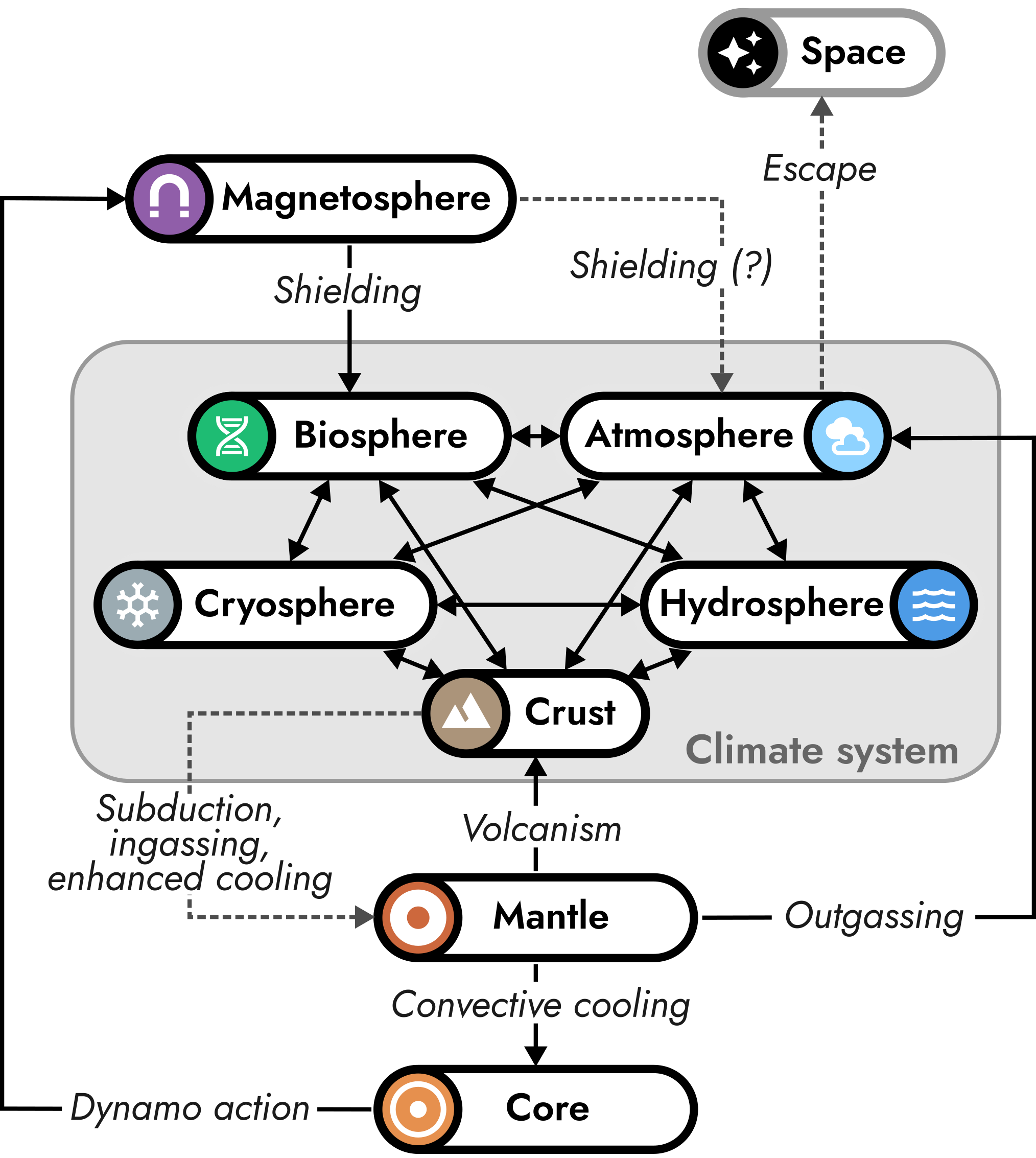}
    \caption{Illustration of the interaction pathways between the interior, surface and atmosphere of a habitable planet. Figure adapted from \citet{Gillmann2022}. Icons are modified from \href{https://pictogrammers.com/}{Pictogrammers}, licensed under the Apache License 2.0 (\url{https://www.apache.org/licenses/LICENSE-2.0}).}
    \label{fig:feedback_diagram}
\end{figure}

These processes form a complex web of interactions which couple the planet's interior to its atmosphere (Fig. \ref{fig:feedback_diagram}). Therefore, the evolution of the atmosphere can not be disentangled from the evolution of the interior. Furthermore, feedback processes arising from these interactions can drastically shape the trajectories a planet's atmosphere may take \citep[see e.g.,][]{noack2012coupling, krissansen2021venus, baumeister2023RedoxState, Gillmann2022}.

The fact that the interior, surface, and atmosphere are linked offers the potential to determine characteristics of exoplanet interiors by observing their atmospheric composition. A prerequisite for this is a comprehensive understanding of these processes, and how they and their interactions manifest themselves in the atmospheric composition.

Furthermore, the presence of climate feedback loops is critical in the context of planetary habitability. To sustain conditions suitable for the presence of liquid water in the long term, the climate must be stable against perturbations in temperature and composition of the atmosphere, for example from volcanic eruptions or the long-term increase in stellar luminosity. This stability is enabled by negative feedback loops, which counteract deviations from the equilibrium state. The most prominent example on Earth is the carbonate-silicate cycle, which regulates the \ce{CO2} content in the atmosphere to a level which permits temperatures suitable for liquid water, thus preventing Earth from being permanently locked into a snowball or hothouse state. On the other hand, positive feedback cycles can exacerbate perturbations, thus acting to destabilize the climate. If these processes are strong enough, they may push a planet irreversibly toward uninhabitable conditions, as may have happened in the case of Venus.
% understanding how they work on exoplanets is therefore important

The interactions and feedback loops of atmosphere, surface, and interior further enable us to distinguish between different types of surfaces based on the atmospheric spectrum of a planet \citep[e.g.,][]{2024Byrne, 2025Herbort}. \citet[][this Topical collection]{Byrne_chapter} reviews our knowledge of Earth's interior, surface, and atmospheric composition through time. They then examine, to the best of our present abilities, the same for Mercury, Venus, Mars, Earth's moon and Jupiter's moon Io, and how they compare to each other. Finally, they consider exoplanetary worlds that may or may not have direct representation within our solar system. 
In the future, it might even be possible to extract water/land ratios of temperate rocky exoplanets by observing their reflected light \citep[see e.g, the discussion in][this Topical collection]{2026Guimondwatervsland}.

\subsection{The carbonate-silicate cycle}

\begin{figure}[ht!]
    \centering
    \includegraphics[width=0.5\linewidth]{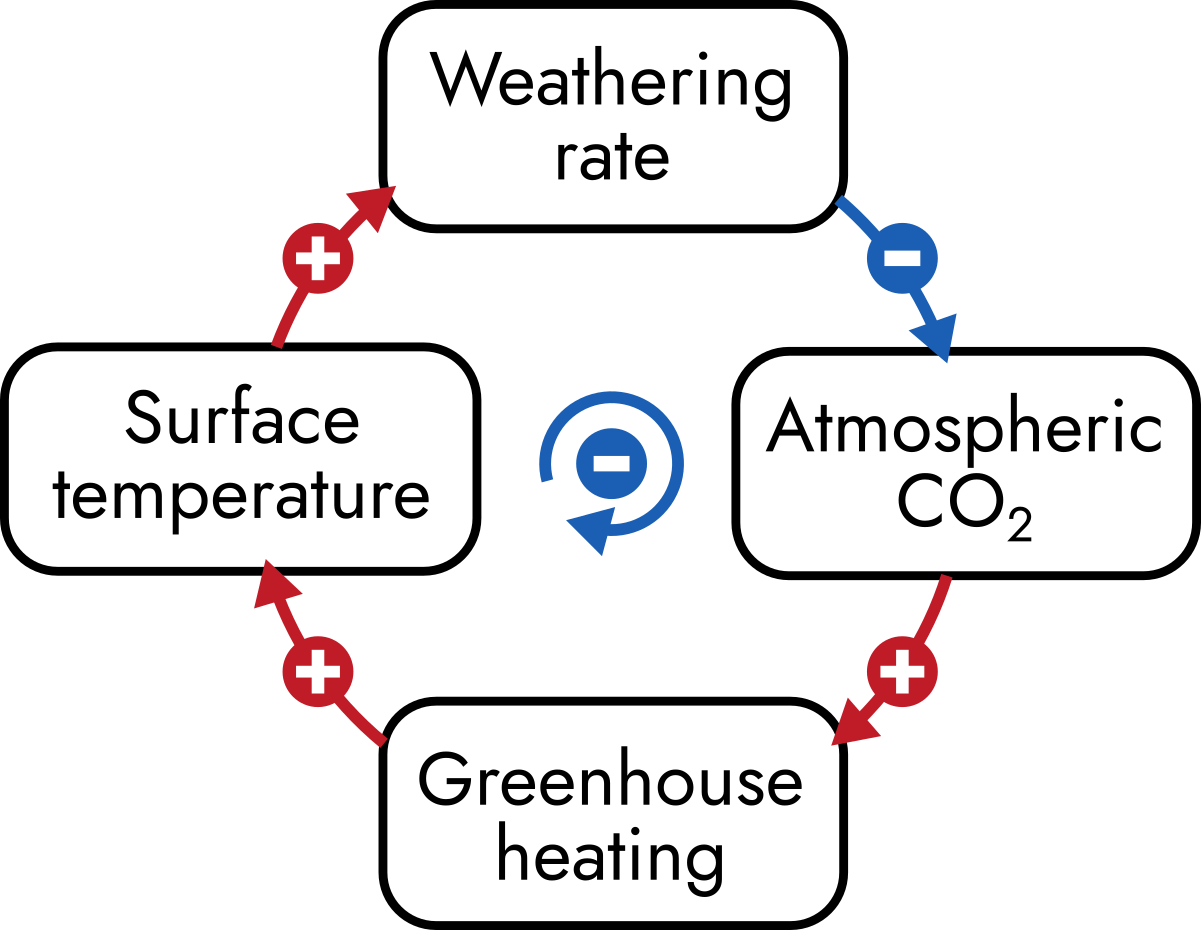}
    \caption{Illustration of the negative feedback loop between \ce{CO2} and silicate weathering that forms the backbone of the carbonate-silicate cycle. Red arrows with a plus mark positive feedbacks. Blue arrows with a minus mark negative feedbacks. After \citet{catling2017AtmosphericEvolution}.}
    \label{fig:feedback_carbon_cycle}
\end{figure}

The carbonate-silicate cycle (Fig. \ref{fig:feedback_carbon_cycle}) is one of the most important long-term feedback cycles on Earth, acting as a ``thermostat'' \citep{catling2017AtmosphericEvolution} for Earth's climate by regulating the amount of \ce{CO2} in the atmosphere \citep{walker1981NegativeFeedback}. Atmospheric \ce{CO2} forms carbonic acid in rain water, which can dissolve silicate minerals (continental weathering). The byproducts are washed into the ocean, where they can form carbonates through biotic and abiotic processes, which then precipitate as sediments on the sea floor. Plate tectonics transports these carbonate layers deep into the mantle. \ce{CO2} is replenished in the atmosphere through volcanic outgassing or metamorphic decarbonation of carbonate rocks at high temperatures and pressures, closing the cycle. Importantly, the weathering processes involve liquid water. As the rate of surface weathering additionally depends on surface temperature and the atmospheric concentration of \ce{CO2}, this forms a negative feedback cycle, stabilizing the surface temperature and climate against perturbations on geological time scales and on a level that permits liquid water. Should the surface temperature fall below the freezing point of water, the water cycle slows down and with it the weathering rates, thus slowing down \ce{CO2} removal from the atmosphere. \ce{CO2} will continue to accumulate in the atmosphere due to volcanic outgassing, which eventually brings the surface temperature above freezing again. Likewise, a positive perturbation in atmospheric \ce{CO2} increases the surface temperature, increasing weathering rates and therefore the draw-down of \ce{CO2}. In short, the carbonate-silicate cycle maintains a \ce{CO2} level which enables liquid water. 

Stagnant-lid planets lack the subducting plates to transport carbonates into the interior. However, as long as fresh rock is being created by volcanic activity, \ce{CO2} weathering cycles can still work even in the absence of plate tectonics. \citet{foley2018CarbonCycling} propose a stagnant-lid carbon cycle where \ce{CO2} is removed by the weathering of newly erupted basaltic rocks. Volcanic activity on stagnant-lid planets tends to be focused locally by hotspots from underlying mantle plumes. Thus, subsequent eruptions bury weathered crust which, through isostatic adjustment, sinks towards the convective mantle, where it may eventually be recycled by lithospheric delamination. With a sufficient level of ongoing volcanism, this process can stabilize the climate on stagnant-lid planets for several billion years \citep{foley2018CarbonCycling, foley2019HabitabilityEarthlike, honing2019CarbonCycling, honing2021EarlyHabitability, baumeister2023RedoxState}.

An important note to make here is that the conventional definition of the habitable zone \citep{kasting1993HabitableZones} implicitly assumes that some form of carbonate-silicate cycle is active to regulate the atmospheric \ce{CO2} concentration. This property of the carbonate-silicate cycle allows a potential population-wide test of the habitable zone. Planets with an active carbonate-silicate cycle near the inner edge of the habitable zone should have low levels of \ce{CO2} in their atmospheres, whereas planets toward the outer edge should have \ce{CO2}-rich atmospheres \citep{catling2018ExoplanetBiosignatures, lehmer2020CarbonatesilicateCycle}. This could be tested statistically by observing several tens of Earth-like exoplanets \citep{lehmer2020CarbonatesilicateCycle}.

\subsection{Water vapor and the runaway greenhouse}

\begin{figure}[ht!]
    \centering
    \includegraphics[width=0.7\linewidth]{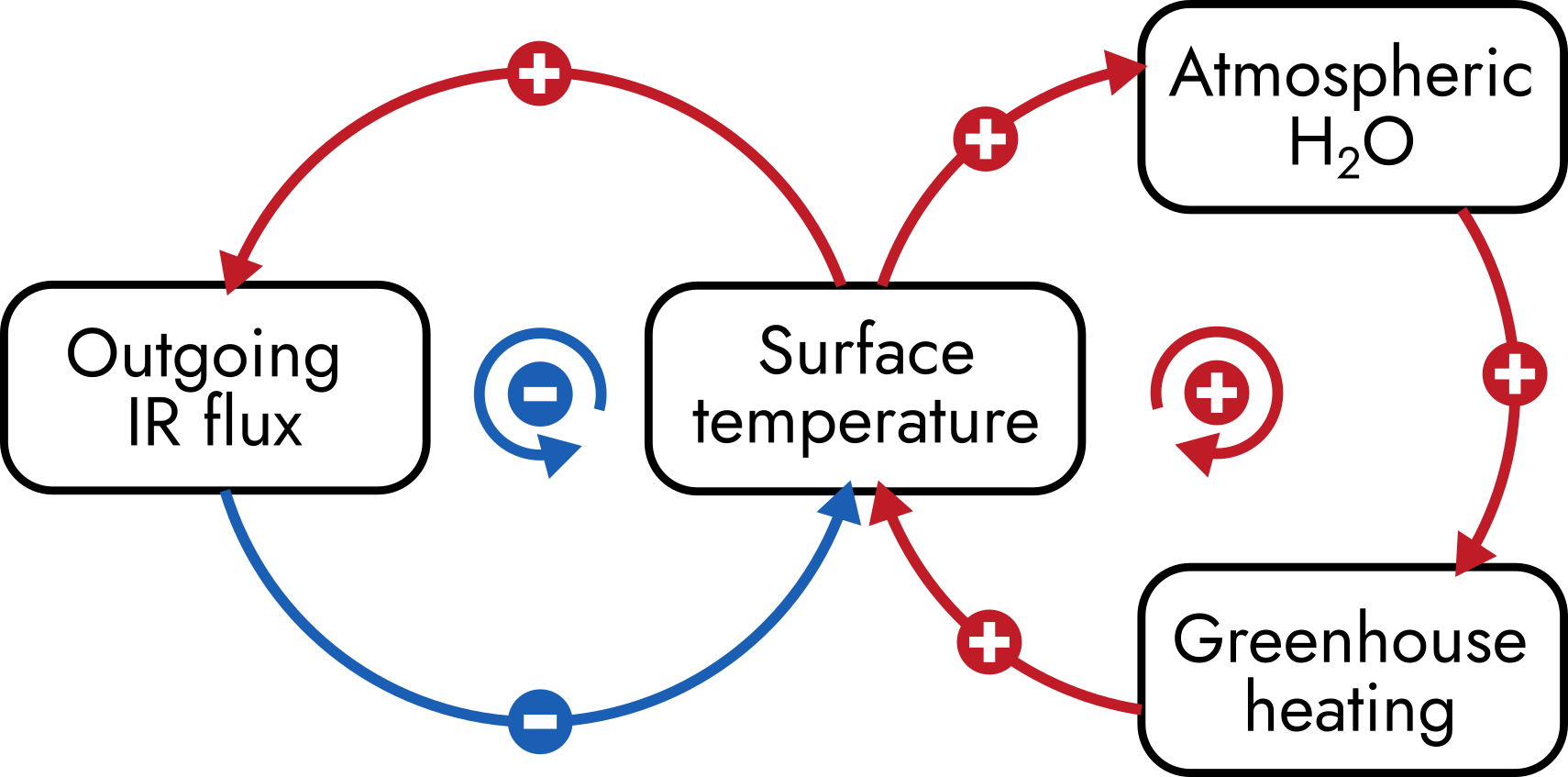}
    \caption{Illustration of the negative feedback loop between surface temperature and outgoing IR flux (left loop) and the positive feedback loop caused by water vapor (right loop). After \citet{catling2017AtmosphericEvolution}.}
    \label{fig:feedback_water_vapor}
\end{figure}

Water plays a central role in many of the climate feedback cycles on Earth, in particular because it is a strong greenhouse gas, but also easily condensable under Earth's temperature and pressure conditions. As a result, water vapor enacts a positive feedback on the climate system (Fig. \ref{fig:feedback_water_vapor}). The concentration of water vapor is a function of surface temperature according to the saturation pressure of water, where higher temperatures increase evaporation as more water vapor can be stored in the atmosphere. Due to the greenhouse effect of water vapor, this in turn raises the surface temperature further. As a result, water acts as an amplifier for climate perturbations. On Earth, this feedback loop is not strong enough to destabilize the climate system, as on short timescales, the surface temperature is moderated by the planet’s outgoing infrared flux \ref{fig:feedback_water_vapor}). As the surface temperature rises, infrared emission increases likewise, thus cooling the planet. 

However, this stabilizing process only works to a certain point. If sufficient water vapor is in the atmosphere, the lower atmosphere becomes optically thick, putting a limit on the outgoing IR radiation flux. If the amount of absorbed stellar irradiation exceeds this critical flux, termed the Komabayashi-Ingersoll limit \citep{komabayasi1967DiscreteEquilibrium, ingersoll1969RunawayGreenhouse}, the surface of the planet heats up uncontrollably in a runaway greenhouse effect, until the entire ocean is evaporated and the surface temperature approaches that of Venus \citep{komabayasi1967DiscreteEquilibrium, ingersoll1969RunawayGreenhouse, kasting1988RunawayMoist, nakajima1992StudyRunaway}. The critical outgoing IR flux puts a limit on the instellation a planet can receive and still retain liquid water on its surface, and is thus used to define the inner edge of the habitable zone. Depending on the study, the runaway greenhouse limit is reached at an effective stellar flux $S_\mathrm{eff}$ (in relation to today's solar flux at Earth's orbit) of 1.4 \citep{kasting1988RunawayMoist} to 1.05 \citep{kopparapu2013HabitableZones} for a planet with an Earth-like mass \citep[for estimations for other surface gravities, see e.g.,][]{pierrehumbert2010PrinciplesPlanetary}.

As soon as the water vapor reaches the stratosphere, photodissociation and subsequent escape of the hydrogen will rapidly desiccate the planet on timescales of tens to hundreds of millions of years \citep[e.g.,][]{kasting1988RunawayMoist} even before a runaway greenhouse is reached. This `moist greenhouse' state \citep{kasting1988RunawayMoist} yields a more conservative estimate on the inner boundary of the habitable zone. Depending on the specifics and complexity of the climate models used, the moist greenhouse limit is somewhere between 1.05 and 1.2 $S_\mathrm{eff}$ \citep{kasting1993HabitableZones, kopparapu2013HabitableZones, leconte2013IncreasedInsolation, gomez-leal2019ClimateSensitivity}.

The runaway greenhouse transition may be detectable in the exoplanet population \citep{schlecker2024BioverseHabitable}. Planets undergoing runaway greenhouse conditions are expected, on average, to exhibit larger radii, which could allow one to map out the inner edge given a sample size of $>100$, for example from the PLATO mission \citep{rauer2025PLATOMission}.

 Strictly speaking, this conventional definition of the habitable zone assumes that the planet is Earth-like, complete with Earth's important geological cycles and abundant surface water. In reality, many factors will influence this limit significantly. For example, \citet{vladilo2013HabitableZone} find that higher atmospheric pressures broadens the habitable zone in both directions. Likewise, desert planets with limited surface water have significantly extended habitable zones \citep{abe2011HabitableZone}. Toward the inner edge of the HZ, the dry stratospheres make these planets stable against a runaway greenhouse, while towards the outer edge, the lack of surface water and clouds available for freezing reduces the ice/albedo feedback (see next section).

\subsection{The ice/albedo feedback and Snowball Earths}
\begin{figure}[ht!]
    \centering
    \includegraphics[width=0.4\linewidth]{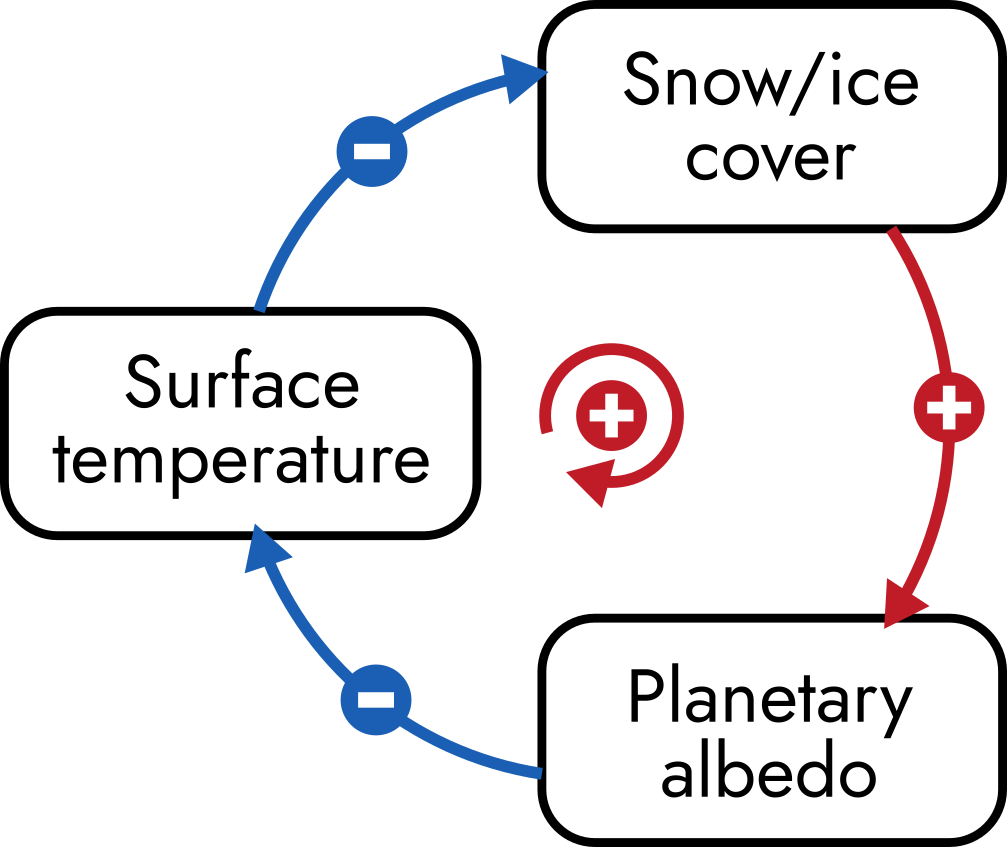}
    \caption{Illustration of the positive ice/albedo feedback loop. After \citet{catling2017AtmosphericEvolution}.}
    \label{fig:feedback_ice_albedo}
\end{figure}

Positive feedback loops can destabilize a planet's climate in both directions. While the water vapor feedback can drive a planet to extreme heating, and even a runaway greenhouse state, the ice-albedo feedback (Fig. \ref{fig:feedback_ice_albedo}) can drive a planet into extreme cooling and global glaciation \citep{budyko1969EffectSolar}. A decrease in temperature enhances snow and ice cover, which increases the planetary albedo, which lowers the absorbed stellar irradiation, thus cooling the surface further and promoting further ice growth. In extreme cases, this may trigger a `Snowball Earth' state, where the surface becomes nearly entirely frozen with little exposed liquid surface water. On Earth, evidence suggests that global glaciation occurred at the end of the Proterozoic around 750\,Myr ago \citep{hoffman1998NeoproterozoicSnowball, hoffman2002SnowballEarth, maruyama2008ModelsSnowball}.

To trigger a Snowball Earth, some process needs to cool the climate sufficiently so that the ice caps extend past $\approx 30^{\circ}$ latitude, at which point the feedback is strong enough to destabilize the climate (assuming a planet similar to Earth). At this point, the ice caps could rapidly grow and envelop the planet, although GCM simulations suggest that liquid water can remain in stable equatorial water belts \citep[see e.g.,][]{hyde2000NeoproterozoicSnowball, abbot2011JormungandGlobal,hoffman2017SnowballEarth,Feulner2023}.

In tidally locked aqua planets around M-dwarfs with low insolations it was initially discovered by \cite{pierrehumbert2010palette} that an ``eyeball" world would exist where there would only be open ocean at the substellar point, but that the planet would otherwise be frozen over. However, their model had the shortcoming that it did not include oceanic heat transport. It was later demonstrated that in fact the pattern of open ocean at the substellar point would more resemble a ``lobster'' shape than the more symmetric/circular ``eyeball'' \citep[e.g.][]{hu2014role,yang2014water,DelGenio2019}. Although the ice-albedo feedback (Fig. \ref{fig:feedback_ice_albedo}) is less pronounced around M-dwarfs given their blackbody radiation peaks more toward the IR where ice albedo is lower than in the visible, it still exists as shown in some studies \citep[e.g.][]{DelGenio2019,Del2019albedos}. Earlier work using 3D GCMs for tidally locked planets around M-dwarfs \citep[e.g.][]{heath1999,Joshi2003} showed that it might be possible to trap most of the planet's water inventory as ice sheets on the night side which led to the idea that these worlds would not be in the habitable zone, but later work using a full complexity model demonstrated otherwise \citep{yang2014water}.
The location of liquid water regions is further influenced by the  existence and distribution of landmass \citep{2022Macdonaldclimate}.

% ocean planets: atmospheres regulate interior through thick water layers (Kitzmann)

\subsection{Clouds and climate}

\begin{figure}
    \centering
    \includegraphics[width=0.8\linewidth]{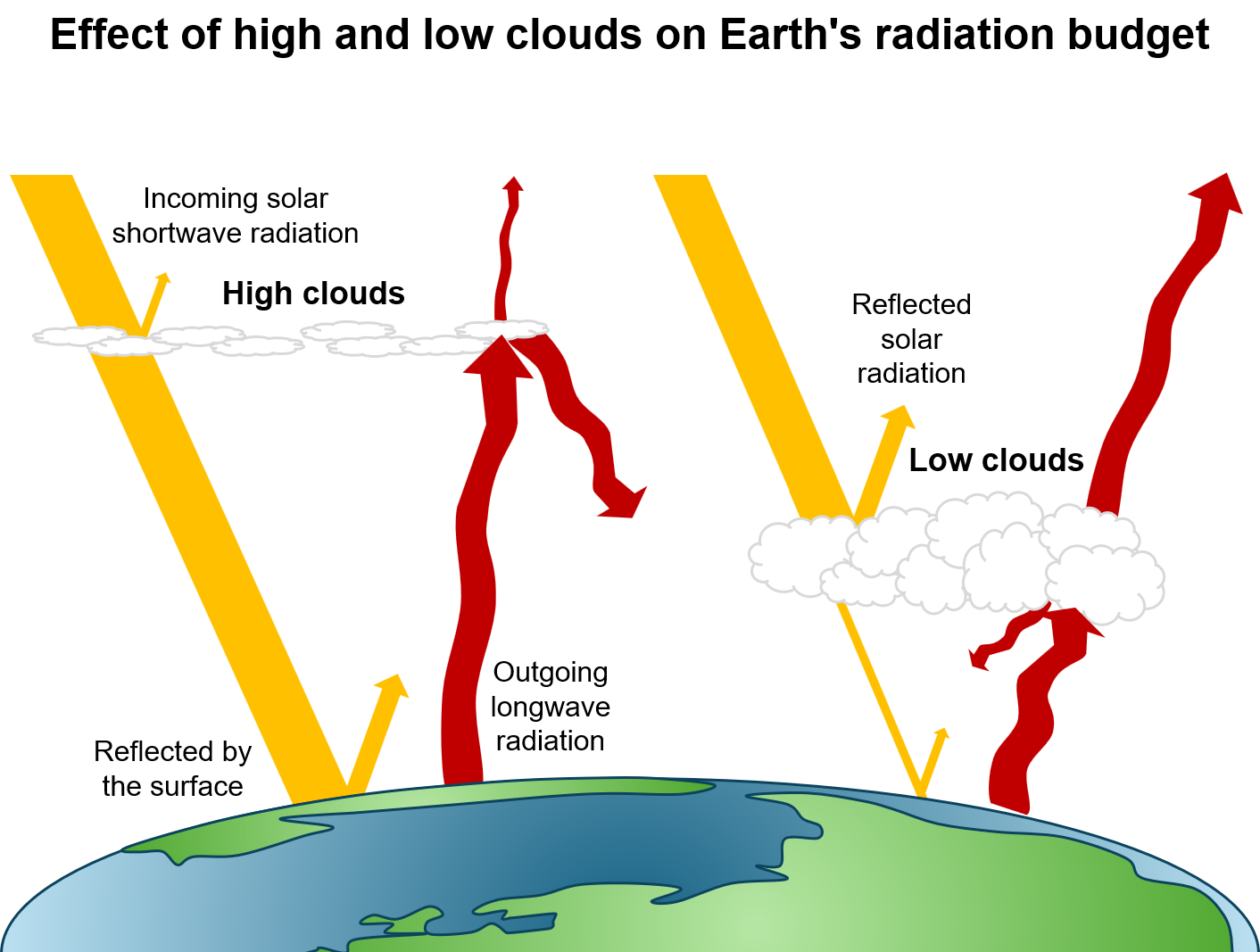}
    \caption{Cloud effects on Earth's radiation budget.}
    \label{fig:clouds}
\end{figure}

Clouds have an extremely important role in the climate of habitable worlds like Earth, but also in other terrestrial worlds in the Solar System such as Venus, Mars and Titan. On Earth, $\sim$67\% of the planet is covered by clouds at any one time with $\sim$55\% coverage over land and $\sim$72\% over the ocean \citep{King2013}. They play multiple first-order roles in the climate such as providing precipitation, but also in the radiative budget. For example, low-level clouds tend to scatter sunlight (short wavelength photons) to space and have little greenhouse effect, whereas high-level clouds tend to trap more long wavelength photons scattered from the surface (See Figure \ref{fig:clouds}) and hence tend to have a greenhouse effect. Clouds have been invoked to solve the Earth's Faint Young Sun Paradox \citep{kasting2010faint,goldblatt2011faint,feulner2012faint,Gudel2015} but different works give conflicting answers  \citep{rondanelli2010can,goldblatt2021earth,goldblatt2011clouds}.
High altitude cirrus type clouds (efficient at trapping longwave radiation) have been proposed as a possible solution for keeping ancient Mars' climate warm enough for surface liquid water \citep[e.g.][]{urata2013simulations,ramirez2017cloud}. However, they did not provide sufficient warming to fully resolve the issue. Venus' thick sulfuric acid clouds not only hide the surface of Venus from visible wavelength observations, but provide an extremely high albedo which has provided a featureless surface for Earth-based observers through the centuries. Until the dawn of the space age it was assumed that the clouds were in fact \ce{H2O} clouds as on Earth, and that beneath the clouds resided a hidden world filled with exotic life \citep[e.g.][]{ORourke2023}. However, \cite{hansen1974interpretation} showed that in fact it was more likely they were 99\% \ce{H2SO4} clouds, as turned out to be true. Clouds on Venus may have also played a highly important role in its early climate history. Work by \cite{Yang2014,Way2016,Way2018,Way2020} showed that for slowly rotating worlds (like Venus) in a temperate state (unlike today's Venus where surface temperatures are $\sim$\qty{450}{\celsius}) a large substellar cloud deck could form. This substellar cloud deck would provide a wide area, high albedo surface reflecting back to space much of the planet's incident stellar radiation. The end result is a world that could have hosted surface liquid water for gigayears, even while its insolation could have ranged from 1.4 to 1.9 times that of modern Earth. This assumes that Venus had a temperate early period, which remains a matter of debate \citep[e.g.][]{Lebrun2013,hamano2013emergence,Turbet2021}.
Venus' thick atmosphere may also have influenced its retrograde rotation rate, but this is still up for discussion \citep{2015Leconteasynchronous,2026FerrazMelloexoplannet}.

The effect of clouds and aerosols on exoplanetary atmospheres and their observables may currently be underestimated, although their importance has been recognised since their first detections (see Section \ref{sec:atmos_exo}) and initial modeling attempts \citep[e.g.,][]{2013_Marley_review}. 

Finally, many 1D climate models do not include clouds. This has important implications not only for the radiative budget, but also in terms of exoplanet atmosphere observations.
Using a full-complexity 3D GCM \cite{fauchez2019impact} has demonstrated that to fully characterize planetary spectra from climate models clouds must be included.

\subsection{Volatile cycles underrepresented in the literature}
%Manuel, Lena, Caroline
\subsubsection{Nitrogen cycle}\label{sec:NCycle}
%\textbf{\todo{Manuel}}

\ce{N2} is often assumed as an unreactive background gas in atmospheres of rocky planets, with its partial pressure, $p$\ce{N2}, remaining constant over geological timescales. This notion, however, does not necessarily hold because of various abiotic and, in the case of the Earth, biotic sources and sinks \citep[e.g.,][]{Stueeken2016,Spross2021}. This is also highlighted by potential swings in the Earth's $p$\ce{N2} and total pressure over time. Whereas $p$\ce{N2} around 3.0 to 3.5 Gyr ago was likely below 1.1\,bar, and potentially as low as 0.5\,bar \citep{Marty2013,Avice2018}, the total pressure of Earth's atmosphere around 2.7\,Gyr ago could have been below $\sim$300\,mbar \citep{Som2016,Rimmer2019Pressure}, out of which a significant fraction was \ce{CO2} \citep{Kanzaki2015CO2}. During the Hadean and early Archean eons, the \ce{N2} partial pressure could have been either around 50\% higher than at present, based on an increased recycling efficiency of subducted \ce{N2} over time \citep{Barry2016,Mallik2018}, or lower, presuming an increase in the \ce{N2} degassing rate during the Archean eon \citep[e.g.,][]{Mikhail2014,Aulbach2016}.

High-energy processes such as lightning \citep{NavarroGonzalez1998,NavarroGonzalez2001}, meteorite impacts \citep[e.g.,][]{Heays2022}, and cosmic rays \citep{TabatabaVakili2016} can provide sufficient energy to break the triple bond of \ce{N2} and to fix it into NH$_x$ and/or NO$_x$. The first two of these processes were more effective on early Earth than at present \citep{NavarroGonzalez1998,Spross2021}, whereas \ce{N2} fixation via cosmic rays and stellar energetic particles (SEP) can be a significant source of \ce{N2} fixation on planets around M dwarfs \citep{Grenfell2020,Scherf2024EEII}. However, SEPs have also been proposed as a means to break up \ce{N2} while creating N$_2$O in early Earth's atmnosphere, possibly helping to solve the Faint Young Sun Hypothesis \citep{airapetian2016prebiotic,Kobayashi2026arXiv}. On Earth, however, biological activity was likely the main cause of fluctuations in the \ce{N2} partial pressure over time \citep{Stueeken2016}. Anaerobic biological nitrogen fixation (BNF), which fixes \ce{N2} into bioavailable forms, originated some time before 3.2 Gyr ago \citep{Stueeken2015,Stueeken2020} and could have significantly depleted Earth's atmospheric \ce{N2} in-line with the low total surface pressure suggested for 2.7 Gyr ago. The first evidence for aerobic biological denitrification was found 2.3 Gyr ago during the Great Oxygenation Event (GOE), marking the onset of Earth's aerobic nitrogen cycle \citep{Zerkle2017b}. Since denitrification releases \ce{N2} back into the atmosphere, it counteracts BNF and has contributed to an increase of $p$\ce{N2} during and after the GOE \citep{Stueeken2016,Spross2021}.

For an abiotic world, the partial pressure of \ce{N2} is expected to increase over geological timescales, at least to a certain extent, as a function of volcanic degassing, nitrogen recycling between the interior and the atmosphere, and abiotic fixation processes \citep{Stueeken2016,Laneuville2018}. The amount of \ce{N2} in the atmosphere can hence allow constraints on surface and interior processes of a planet. As long as strong atmospheric losses can be ruled out, little \ce{N2} in the atmosphere likely implies large abiotic fixation processes, whereas large amounts of atmospheric \ce{N2} hint towards a lack of plate tectonics and/or a hydrological cycle, which both aid in recycling \ce{N2} back into the mantle \citep{Laneuville2018}. However, to significantly deplete an \ce{N2} atmosphere via abiotic fixation processes on a planet with an anoxic atmosphere in contact with liquid water, it must be $>$100 times larger than Earth's present-day abiotic fixation rate \citep{Hu2019N2}.

For Mars, isotopic signatures imply that most of its \ce{N2} was lost into space early in its history \citep{Kurokawa2018,Hu2022Mars}, leaving little traces on how the Martian geological evolution may have shaped its nitrogen cycle. For Venus, it is interesting to note that its atmosphere hosts about 3\,bar of \ce{N2}. This is roughly the same amount of nitrogen presently assumed to be in Earth's bulk silicate mantle, crust, and atmosphere \citep{Marty2012,Li2024N2}, which suggests a well-degassed Venusian interior. However, $^{40}$Ar abundances imply that Venus is less degassed than Earth \citep{Kaula1999,ORourke2015}. This is a mystery to be solved by future missions to Venus, but it may point toward a comparatively early degassing of (most of) its atmosphere and a subsequent lack of plate tectonics. During the magma ocean phase, or at a later point in its history, a runaway greenhouse phase on Venus led to the escape of hydrogen after photolysis of \ce{H2O} and a likely increase in the mantle's oxidation state via the left-behind oxygen \citep[e.g.,][]{Gillmann2022}. Since the degassing capacity of \ce{N2} increases towards larger surface pressures \citep{gaillard2014theoretical} and higher oxygen fugacities \citep{Gaillard2022}, this indicates an efficient degassing of \ce{N2} into Venus' atmosphere \citep{Wordsworth2016,Gaillard2022}, even during its present-day conditions \citep{Gillmann2022}. Little \ce{N2} fixation and inefficient recycling due to the lack of plate tectonics and/or a hydrological cycle during the rest of Venus' history kept most of the \ce{N2} in its atmosphere, compatible with the general abiotic nitrogen cycle scenarios outlined by \citet{Laneuville2018}.

Finally, our solar system hosts three additional bodies with \ce{N2}-dominated atmospheres, i.e., Titan, Pluto, and Triton. All of these are icy bodies with large nitrogen reservoirs that were either accreted from cometary ammonia ices, complex refractory organics and/or directly from \ce{N2} ices in the protosolar nebula \citep[e.g.,][]{Scherf2020}. Titan hosts a thick \ce{N2}-dominated atmosphere, which likely originated through the decomposition and degassing of ammonia ices and complex organics from its interior \citep{Glein2015,Erkaev2021}. Pluto and Triton, on the other hand, host tenuous \ce{N2}-dominated atmospheres in evaporation pressure equilibrium with their surface, whose characteristics and evolution are poorly understood \citep[e.g.,][]{Scherf2025}. Their location in the outer solar system, their overall composition, and large accreted nitrogen reservoirs lead to completely different mechanisms dominating their nitrogen cycling compared to Venus, Earth, and Mars. However, planetary migration in exoplanetary systems could move such water- and nitrogen-rich bodies closer to their host stars, where they may end up as exotic ocean worlds with highly reducing \ce{CO2} or \ce{N2}-dominated atmospheres, at least as long as their planetary mass is sufficiently high to hinder the hydrodynamic escape of nitrogen \citep{Spross2021}. The gradual warming of a host star at the end of its main-sequence lifetime could have a similar effect; Titan, for instance, could enter a brief phase of liquid water-surface habitability in roughly 6\,Gyr from now during the Sun's red giant phase \citep{Lorenz1997,sparrman2024multiple}. Such a scenario might in principle be distinguishable from a rocky exoplanet with a deep global ocean, since \ce{N2} degassing, and hence the build-up of a thick \ce{N2}-dominated atmosphere, might be inhibited on water worlds \citep{KrissansenTotton2021b}. 

\subsubsection{Sulfur cycle}
On modern Earth, sulfur cycles between the hydrosphere, atmosphere, biosphere, pedosphere, and upper lithosphere through geological, biological, and anthropogenic processes. At the same time a large, poorly constrained amount remains sequestered in the core and lower mantle, excluded from the active global cycle \citep{schoonen2016sulfur}. The largest reservoir is the upper lithosphere and marine sediments, which contains an estimated $10^{12}$ g of sulfur stored as gypsum, anhydrite, metal sulfides, or elemental sulfur \citep{brimblecombe2003global}. From the lithosphere, sulfur can enter the atmosphere through volcanic degassing, but the atmosphere itself represents the smallest reservoir, holding sulfur primarily as gases such as \ce{H2S}, \ce{SO2}, DMS, and OCS, or as aerosols like \ce{H2SO4} and (NH$_4$)$_2$SO$_4$; additional inputs include dust particles, biogenic emissions, and biomass burning \citep{schoonen2016sulfur}. These atmospheric sulfur species are rapidly removed by precipitation and deposition, transferring sulfur into the hydrosphere and pedosphere. The hydrosphere, particularly the oceans, constitutes the second-largest reservoir and the site of the most active cycling, with sulfur present mainly as sulfate \citep{brimblecombe2003global}. Inputs to this pool include weathering of sulfide- and sulfate-bearing rocks and atmospheric deposition, while losses occur through evaporation and microbial processes that reduce sulfate to sulfide or convert it into organo-sulfur compounds. In the biosphere and pedosphere, sulfur is found in both organic forms, such as amino acids, and inorganic forms, including sulfates, sulfides, and elemental sulfur \citep{brimblecombe2003global}. Today, human activities strongly influence the sulfur cycle by releasing large amounts of \ce{SO2} into the atmosphere and hydrosphere through the burning of sulfur-rich fossil fuels, with estimates suggesting that more than half of the sulfur transported to the oceans by rivers and about 75\% of atmospheric sulfur emissions are anthropogenic in origin \citep{berner2012global}.

Venus' atmosphere today contains about 200 ppm sulfur, primarily in the form of \ce{SO2} and its oxidation product \ce{H2SO4} \citep{Basilevsky2003}. During planetary accretion and any early magma ocean stage, volatiles dissolved in the melt---including S, C, and H---would have been outgassed into the nascent atmosphere, and models suggest that this large-scale degassing delivered substantial sulfur, both as \ce{SO2} and reduced sulfur gases, very early in Venus’s history \citep{Fegley2003}. After the magma ocean cooled, continuing volcanic activity provided the dominant source of sulfur to the atmosphere \citep{Fegley2003}. Geological evidence indicates that Venus underwent major resurfacing episodes, as reflected in its relatively young surface age and widespread lava plains. If resurfacing occurred in pulses, each episode would have injected large \ce{SO2} loads into the atmosphere, with significant consequences for cloud formation and atmospheric chemistry \citep{Taylor2009}. In the present atmosphere, intense ultraviolet radiation drives photolysis of \ce{CO2} into CO and O, and the resulting oxygen oxidizes \ce{SO2} to \ce{SO3}, which ultimately hydrates to form \ce{H2SO4}. Sulfuric acid then condenses into droplets and aerosols, forming the planet's extensive cloud decks \citep{Marcq2018}. \ce{SO2} also dissolves into cloud droplets, where interactions with salts or hydroxides can sequester sulfur temporarily from the gas phase \citep{Rimmer2023}. Unlike Earth, however, Venus lacks a rainout mechanism to remove \ce{H2SO4} from the atmosphere. Instead, laboratory experiments suggest that \ce{SO2} can react directly with surface basalts to form secondary sulfates such as \ce{CaSO4}, \ce{MgSO4}, and \ce{Na2SO4}, providing a significant near-surface sulfur sink \citep{RadomanShaw2022, Renggli2019}. Because Venus does not exhibit Earth-like plate tectonics, recycling of sulfur into the mantle is unlikely \citep{Byrne2021, Gulcher2020}. Nevertheless, high resurfacing rates could permit burial and partial remobilization of sulfur-bearing basalts into magmatic systems, offering a possible, if incomplete, mechanism for closing the sulfur cycle.

% Lena, Caroline
Io, the innermost of Jupiter’s Galilean moons, is the most volcanically active body in the solar system due to intense tidal heating driven by the gravitational interaction with Jupiter and its neighboring moons, particularly Europa \citep{geissler2008galileo}. During its early evolution, Io likely accreted volatiles such as hydrogen, carbon, and nitrogen, but these were lost over time as a result of its vigorous volcanism and interactions with Jupiter’s magnetosphere. In particular, elastic scattering by high-energy protons in the magnetosphere caused the ejection of these lighter, reduced gases---such as \ce{H2} and CO---from Io's exosphere into the surrounding plasma environment, gradually oxidizing the moon’s surface chemistry and leaving sulfur predominantly in the form of \ce{SO2} \citep{pollack1980origin}. Unlike the lighter volatiles, sulfur is heavier and more likely to remain near the surface, where it condenses and precipitates, leading to Io’s desiccation and making it an archetype for a planetary body governed by sulfur-driven geochemical processes. Sulfur gases such as \ce{SO2}, \ce{S2}, and SO are either directly degassed from Io’s volcanoes or produced through interactions between lava and surface \ce{SO2} ice. These volcanic emissions often condense to form various sulfur ices, giving Io its characteristic multicolored surface with hues of white, yellow, orange, red, green, and black. Some volcanic plumes can reach altitudes of up to 500\,km, and gases that escape Io’s gravity contribute to a surrounding neutral torus. In this region, sulfur species are ionized through photochemical reactions to form the Io plasma torus, which primarily consists of ions such as \ce{S+}, \ce{O+}, \ce{SO+}, and \ce{SO2+} \citep{thomas2004io}. These ions can then be accelerated within Jupiter’s magnetosphere and transported to Jupiter’s atmosphere or to other Galilean moons \citep{lodders2024sulfur}. The ongoing degassing of volcanic vents, along with sublimation and sputtering of surface ices, sustains Io’s tenuous atmosphere, which has a surface pressure ranging from 0.3 to 3\,nbar on the day side and consists mostly of \ce{SO2} (90\%) along with minor constituents like \ce{S2}, \ce{S2O}, NaCl, and KCl, under average daytime temperatures around 130\,K \citep{spencer2000io}. Additionally, \ce{SO2} ice on the surface undergoes radiolysis, forming \ce{SO3}, and this ice is eventually buried and recycled back into Io’s interior, where it is re-incorporated into volcanic activity—thereby closing Io’s sulfur cycle.

Sulfur cycles as observed on Io have recently gained increased attention, due to tentative detections of \ce{SO2} in the atmosphere of rocky planets \citep{bello2025evidence,nicholls2025volatile}.

\section{Effect of life on the atmosphere}\label{subsec:Life_Atmosphere}
\subsection{Habitable atmospheres}
%Leads: Lena + ?, Mike, Manuel
% I split this up into first habitable atmospheres (which I moved to the beginning of the section), then inhabited atmospheres 

%\lena{
While the composition of the modern atmosphere of Earth is clearly linked to the existence of life (see above and in Figs.~\ref{fig:atms_evol},\ref{fig:LUVOIR}), a habitable atmosphere does not necessarily need to be comprised of oxygen and nitrogen. As a matter of fact, since oxygen reacts easily and can therefore be very destructive, life first needed to adapt to the current oxygen-rich conditions before thriving in it (because of its reactive potential, it serves as an excellent energy source for life). While the exact atmospheric composition and redox state of the Hadean and Archean Earth remain unknown, we do know that free oxygen was mostly found at levels $<$10$^{-6}$ Present Atmospheric Level prior to the Great Oxygenation Event (explained in more detail in Section \ref{ssec:inhabited}). Such reducing conditions are favorable for prebiotic chemistry \citep[e.g.,][]{Luisi2016,Lingam2020Oxygen}, as demonstrated in the Urey-Miller experiment and follow-up experiments, to spontaneously create amino acids and other prebiotic molecules in reducing to moderately oxidizing atmospheres \citep{miller1959organic}. 
Macroscopic life, on the other hand, may be impossible without incorporating oxygen, and past bursts in growth of individual life forms are directly correlated with atmospheric oxygen levels \citep{payne2009two, vilovic2023variations}. We therefore need to distinguish between different habitable conditions — for the origin of life, as well as for the long-term evolution of macroscopic life. This implies, specifically, that an abiotically produced oxygen-rich atmosphere may not be favorable for the origin of life.
%}

The atmosphere composition suitable for macroscopic, complex animal-like life has been suggested to lie in relatively narrow pressure limits for \ce{O2} \citep{Catling2005}, \ce{CO2} \citep{Schwieterman2019}, and \ce{N2} \citep{Ramirez2018}; see Fig.~4 in \citet{Lammer2024EEI} for a review.  
These are mainly toxicity limits for \ce{CO2} and \ce{N2} that may, in principle, vary for other inhabited planets based on the specific evolutionary pathways taken \citep{Scherf2024EEII}. \ce{O}, on the other hand, provides physical limits that can be expected to be valid for any macroscopic animal-like life \citep{Catling2005}. Oxygen provides by far the highest free energy of all elements that can considerably accumulate in an atmosphere, and aerobic metabolism supplies about an order of magnitude more energy than anaerobic metabolism \citep[see in-depth discussion in][]{Catling2005}. Anaerobic organisms are therefore highly restricted in their growth and size, whereas aerobic macroscopic life is either limited in its physical size by oxygen diffusion (and hence \ce{O2} partial pressure), or has to evolve a circulatory system that transports oxygen effectively through the entire body. An \ce{O2} mixing ratio beyond $\sim$30\%, on the other hand, leads to uncontrollable flammability, \citep[e.g.,][]{Belcher2010,balbi2024oxygen} potentially serving as a regulatory factor for the \ce{O2} mixing ratio on an inhabited planet \citep{Belcher2010,Belcher2021}. Therefore, we need to keep in mind that the variety of habitable atmospheres might be much more restricted for macroscopic than for microscopic life. In addition, we highlight that ``habitable'' does not necessarily mean ``inhabited'' \citep[see review by][]{Cockell2016}.

\subsection{Inhabited atmospheres like Earth}
\label{ssec:inhabited}
\begin{figure}
    \centering
    \includegraphics[width=0.8\linewidth]{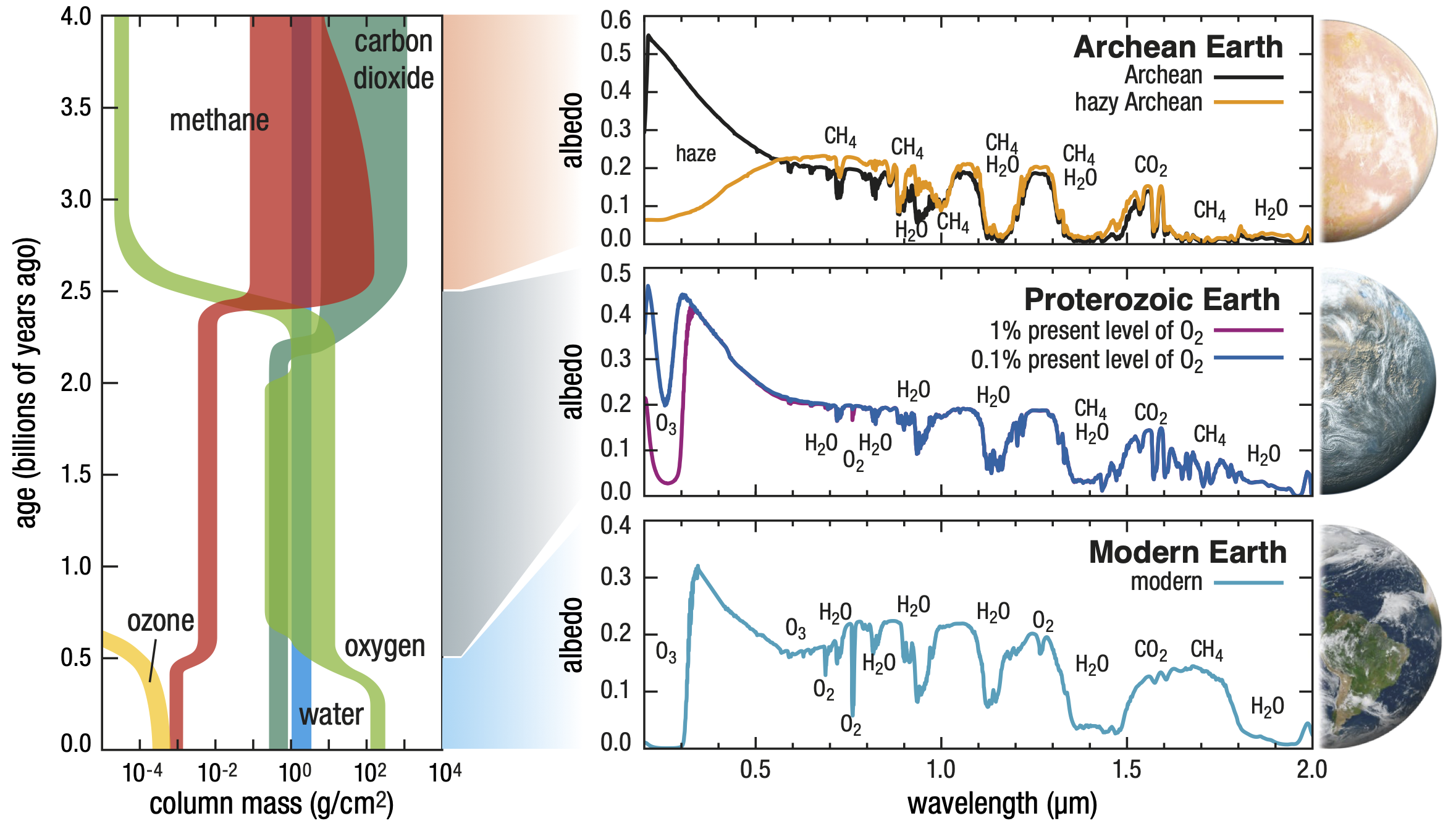}
    \caption{Hypothetical transmission spectra of Earth through time as seen with the proposed LUVOIR/HWO telescope \citep{LUVOIR2019}.  Credit: G. Arney, S. Domagal-Goldman, T. B. Griswold (NASA GSFC)}
    \label{fig:LUVOIR}
\end{figure}

%\MJW{
We start first with what we know best, Earth. Although the dates for the rise of life on Earth are debated \citep[e.g.][Section 2.2.3]{Westall2023}, it is clear that some of the first life forms to influence the atmosphere were methanogens in the Archean (3-8--2.5\,Ga), which used \ce{H2} and \ce{CO2} to produce \ce{CH4} via the reaction \ce{CO2}+\ce{4H2}$\rightarrow$\ce{CH4}+2\ce{H2O}. Given the lack of \ce{O2} in this epoch, the longer residence lifetime of \ce{CH4} (compared to today) would imply substantial build-up of this powerful greenhouse gas. It has long been speculated that it could have played a role in the Faint Young Sun Paradox given that Earth's insolation was 20--25\% less in the Archean than today, and a modern atmospheric composition at that time would otherwise produce a snowball Earth that is not supported by proxy data \citep[e.g.][]{Wilde2001,Valley2002}. At the same time, it is clear that \ce{CO2} partial pressures were also higher \citep[e.g.][]{Feulner2023}. However, one cannot have too much of a good thing, if the \ce{CH4} to \ce{CO2} ratio is $\sim$0.1 or higher hydrocarbon hazes (e.g., \ce{C5H4}, \ce{C4H2}, etc.) can form, which exert a cooling effect on the atmosphere \citep[e.g.][]{Haqq2008,Arney2018}. \ce{CH4} in particular may be observable in planetary transmission spectra from JWST or LUVOIR/HWO (see Figure \ref{fig:LUVOIR}), but clouds and hydrocarbon hazes make the detection of \ce{CH4} and other species more challenging \citep[e.g.][]{fauchez2019impact}.
However, \ce{CH4} may also be produced abiotically via processes in the upper mantle and crust \cite[e.g.][]{DesMarais2002}, motivating the need to rule out the detection of \ce{CO} --- a so-called anti-biosignature --- to strengthen the case for \ce{CH4} as a biosignature \citep{2022Krissansen-Tottonunderstanding,2022Thompsoncase}.

As we move forward in Earth's history into the Proterozoic (2.5\,Ga--541\,Ma) at around 2.5\,Ga we have the Great Oxygenation Event (GOE) which is believed to have been driven by the rise of cyanobacteria \citep[e.g.][]{Kasting_Catling2003,2014Lyons_oxygen}. For many years it was assumed that \ce{O2} was an excellent biosignature \citep[e.g.][]{DesMarais2002}, but it was later realized that it is in fact an ambiguous one. For example, it is possible to create large quantities of \ce{O2} abiotically via photolysis of \ce{H2O} coupled with high rates of atmospheric escape producing 100s if not 1000s of bars of \ce{O2} possibly in a runaway or post-runaway greenhouse state \cite[e.g.][]{luger2015extreme,Meadows2017}. But there are other mechanisms for producing \ce{O2} and even \ce{O3} abiotically \citep[e.g.][]{Schwieterman2018}. \cite{Schwieterman2018} provides an in-depth review of exoplanet biosignature gases (also see Section~\ref{sec:photo}). 
Other work by \cite{leger2011presence} showed that abiotic build-up of \ce{O2} on a habitable planet was unlikely, but later work by \cite{Narita2015} maintained that it was still possible.
Earth provides three main epochs in which its atmosphere has changed sufficiently to be distinguishable if observed as an exoplanet, in the Archean, Proterozoic, and Modern epochs as can be seen in Figure \ref{fig:LUVOIR}.

%}

%\todo{\textbf{Lena: add paragraph on why we don't have a section of biosignature}} removed based on comments from co-authors

\section{Role of photochemistry}
\label{sec:photo}
\begin{figure}
    \centering
    \includegraphics[width=0.8\linewidth]{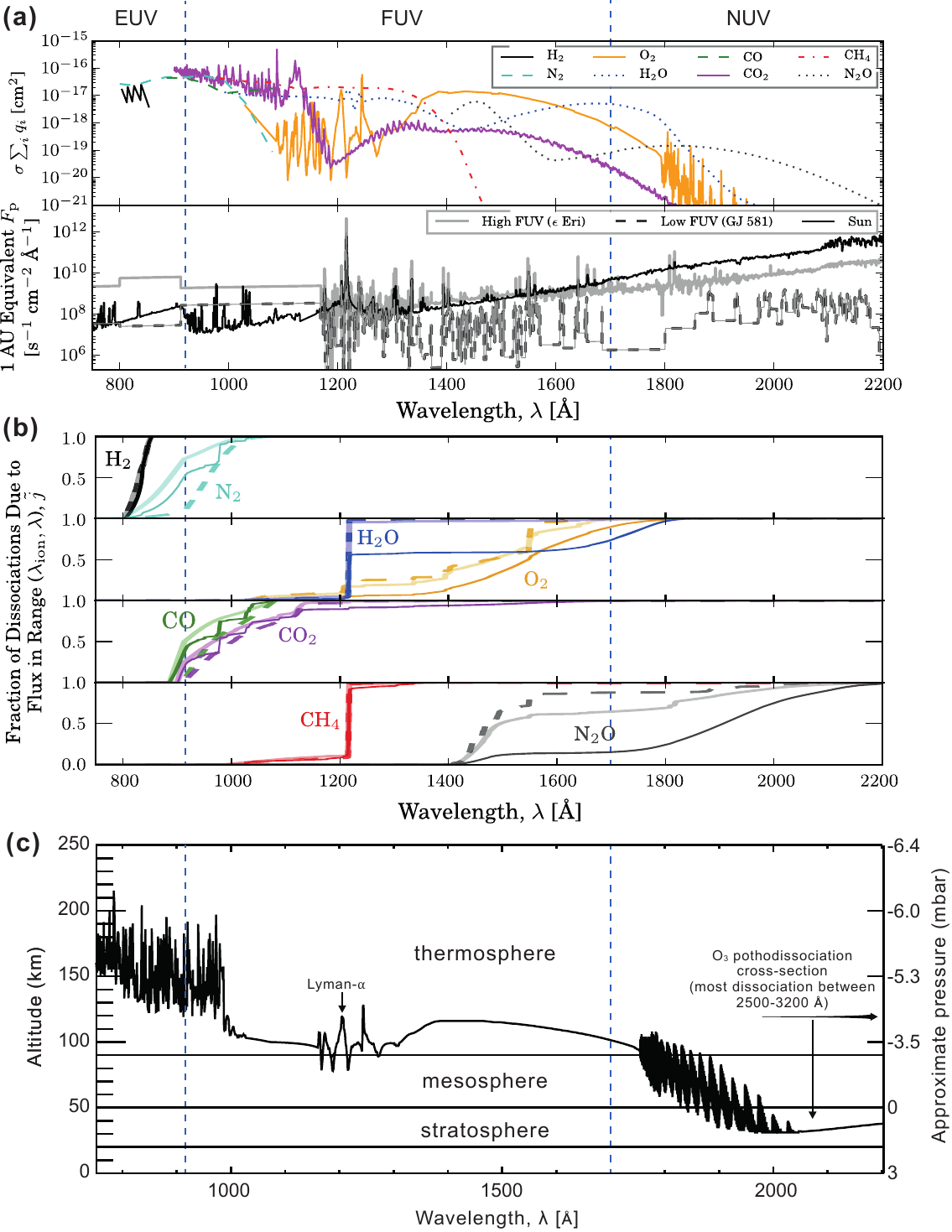}
    \caption{a) top panel: photodissociation cross-sections of various molecules. bottom panel: spectral energy distribution of the Sun, the most active ($\epsilon$ Eri; with $\sim$0.85\,M$_{\odot}$ and an age of $\sim660$\,Myr) and the least active (GJ 581; with $\sim$0.3\,M$_{\odot}$ and an age of $\sim9.5$\,Gyr) star of the MUSCLES survey \citep{France2016,Loyd2016} defined by the ratio of the FUV(far ultraviolet) to bolometric flux. The panel shows the photon flux density scaled to a bolometric flux equivalent to Earth's insolation. (b) Cumulative photodissociation spectra of the same molecules based on the curves in panel (a). The three curves per molecule correspond to the three aforementioned reference stars. (c) Absorption altitudes and approximate pressure levels of unit optical depth (i.e., about 63\% of the flux is absorbed above this altitude) of the same wavelength range as depicted in panels (a) and (b) for the Earth's present atmosphere; note that the absorption altitudes differ for different atmospheric compositions. For the absorption altitudes, see, e.g., \citet{Meier1997} and \citet{Woods2000}. This panel also schematically depicts the wavelengths at which \ce{O3} photodissociates. However, most of the photodissociation of \ce{O3} for the Sun and $\epsilon$\,Eri takes place between 2500 and 3200\,{\AA} \citep[][see also their Fig.~10]{Loyd2016}. The vertical blue dashed lines depict the EUV (extreme ultraviolet), FUV (far ultraviolet), and NUV (near ultraviolet) wavelength ranges. Subfigures (a) and (b) are adopted from \citet{Loyd2016}; the data of the absorption altitudes in (c) is taken from \citet{Woods2000}, their Fig.~2.}
    \label{fig:photochem}
\end{figure}

% Kanako Seki
The incident near-UV to X-ray flux from the host star is an important driver of chemistry in planetary atmospheres, responsible for various photodissociation and photoionization reactions \citep[e.g.,][]{Yung1999Photochemistry,Bauer2004}. The far-UV (920-1700\,{\AA}) is absorbed at pressure levels above about 0.1\,mbar, that is, in the mesosphere and the lower thermosphere in the case of the Earth, whereas the near-UV (1700-3200\,{\AA}) can reach further down to the stratosphere and troposphere (see Figure~\ref{fig:photochem}, panel c). Various molecules such as \ce{CH4}, \ce{O2}, \ce{H2O}, \ce{CO2}, and \ce{N2}O have large parts of their photodissociation cross-sections in the far-UV, with the latter three of them expanding toward the near-UV \citep[e.g.,][]{Loyd2016,Linsky2025}, as can be seen in panels (a) and (b) of Figure~\ref{fig:photochem}. These UV photons drive oxygen photochemistry via the dissociation of molecules such as \ce{H2O}, \ce{O2} and \ce{CO2}, and subsequent three-body reactions, for example, $\rm O + \ce{O2} + M \rightarrow \ce{O3} + M$, where M is an atmospheric molecule needed to ensure momentum balance, can then further produce ozone \citep[e.g.,][]{Bauer2004}. Low-mass M dwarfs are specifically active in the far-UV \citep[e.g.,][]{France2016,Loyd2016} and it was suggested that habitable zone planets orbiting these stars can therefore build up atmospheres with high \ce{O2} and \ce{O3} mixing ratios produced purely by photochemistry \citep[e.g.,][]{Gao2015Phot,Harman2015}. More recent studies indicate that these results are sensitive to various model assumptions, such as the implemented lightning rates and the related production of NO$_{\rm x}$ species \citep{Harman2018}, the chosen photodissociation cross-sections \citep{Ranjan2020}, and the pressure range of the upper model boundary \citep{Ranjan2023Oxygen}. The abiotic build-up of photochemically produced \ce{O2} may hence be less efficient as recently assumed, since CO can efficiently recombine to \ce{CO2} via OH as a catalyst under properly chosen model conditions \citep{Harman2018,Ranjan2023Oxygen}. Still, \citet{Ranjan2023Oxygen} point out that an abiotic build-up of \ce{O3} remains feasible in these photochemical models around M dwarfs.

However, photochemical models often do not include the thermosphere, that is, the region in the atmosphere above the mesosphere where the XUV flux -- the extreme-UV (100-920\,{\AA}) and the X-ray part of the spectrum (10-100\,{\AA}) -- is predominantly absorbed. In the Earth's atmosphere, this happens at a relatively broad range of pressure levels between $\sim0.1\,$mbar in the upper mesosphere and $\sim10^{-6}\,$mbar in the thermosphere \citep[e.g.,][]{Bauer2004}. The absorbed XUV flux (or, more broadly, the absorbed wavelength range below the Lyman-$\alpha$ line at 1215\,{\AA}) in the upper atmosphere is an important driver for photodissociation of, e.g., \ce{H2}, CO, \ce{N2}, and \ce{CO2} \citep{Loyd2016,Linsky2025}, as can be seen in panel (b) of Figure~\ref{fig:photochem}. In addition, it is primarily responsible for photoionization, the corresponding formation of an ionosphere, and thermospheric heating \citep{Bauer2004}. Because of the latter, highly irradiated atmospheres, such as planets in the HZ of M dwarfs \citep[e.g.,][]{VanLooveren2024Trappist,Scherf2024EEII}, can have significantly expanded atmospheres and strong thermal escape due to the large amounts of energy deposited into the thermosphere. High-energy photons in the thermosphere can also dissociate molecules that cool the atmosphere, such as \ce{CO2} \citep{Tian2009,Johnstone2021Earth}. As the altitude at which the XUV flux is typically absorbed increases for expanding atmospheres, it is important to note that thermal escape calculated via the energy-limited escape formulation (see Equation~\ref{eq:en-lim}) depends on $R_{\rm XUV}^2R_{\rm pl}$, and not, as often simplified, on $R_{\rm pl}^3$ \citep[e.g.,][]{2016Erkaevmassloss}. Estimating thermal losses via the latter simplification can, therefore, lead to a significant underestimation of atmospheric escape rates.

Since the XUV flux absorbed in the thermosphere drives photodissociation, photoionization, and atmospheric escape, one has to take account of the upper atmosphere (i.e., the thermosphere above $\sim$80\,km) for evaluating whether \ce{O2} and \ce{O3} may constitute a false-positive biosignature \citep[e.g.,][]{Meadows2018,Schwieterman2024}. As thermal escape preferentially removes the lighter species from the atmosphere, the photodissociation of \ce{H2O}, mostly via absorption at the Lyman-$\alpha$ line (see Figure~\ref{fig:photochem}), can lead to an accumulation of \ce{O2} compared to H \citep{Tian2014O2,luger2015extreme,wordsworth2014abiotic,Wordsworth2018,Johnstone2020Water}. The high XUV activity of M dwarfs their habitable zones \citep[e.g.,][]{Johnstone2021Stars} favors such scenarios, but highly active young K and G stars can also lead to an accumulation of \ce{O2} on planets within their HZs \citep{Johnstone2020Water}. In addition, one should also note that CO, which drives the recombination of \ce{CO2} and hence the decrease of abiotically produced \ce{O2} \citep{Ranjan2023Oxygen}, is dissociated in the thermosphere as most of its photodissociation cross-section is in the XUV \citep{Loyd2016}.

Another relevant photochemical process in the upper atmosphere is the production of suprathermal atoms. This is an important source of non-thermal escape in terrestrial atmospheres (see Section~\ref{sec:solarsystem}, specifically for Mars) since the related photochemical reactions usually include the excitation of atoms above energy levels needed for gaining escape velocity \citep{Bauer2004}. These reactions comprise radiative recombination reactions, which can be expressed as
\begin{equation}\label{eq:radRec1}
  \rm X^+ + e \rightarrow X^{\ast} + \Delta E,
\end{equation}
where X is the initially ionized species, e is a photoelectron, and $\ast$ denotes the excited state of species X. They also include dissociative recombination reactions in the form of
\begin{equation}\label{eq:radRec2}
  \rm XY^+ + e \rightarrow XY^{\ast} \rightarrow X^{\ast} + Y + \Delta E,
\end{equation}
where XY is the molecule, X and Y are its dissociation products, and $+$ and $\ast$ again refer to the ionized and excited states, respectively. In both reaction pathways, the excess energy $\Delta$E goes into the kinetic energy of the neutralized particles \citep{Bauer2004}. Dissociative recombination coefficients (in the order of 10$^{-7}$\,cm$^3$\,s$^{-1}$) are typically orders of magnitude larger than radiative recombination coefficients (in the order of 10$^{-12}$\,cm$^3$\,s$^{-1}$), making dissociative recombination one of the most important photochemical loss processes in planetary atmospheres \citep[e.g.,][]{Biondi1969,Bauer2004}; see Section~\ref{sec:solarsystem} for a discussion on the role of photochemical escape on Venus and Mars.

Photochemistry in the upper atmosphere also depends on atmospheric composition. For hydrogen-rich atmospheres, lower atmospheric layers are dominated by molecules such as \ce{H2}, \ce{H2O}, and \ce{CH4} \citep[e.g.,][]{Hu2012Photo,Madhusudhan2016}. There, the radical OH is much less abundant by orders of magnitude than atomic H. OH, which forms from the dissociation of \ce{H2O}, quickly reacts with \ce{H2} to reform \ce{H2O}, a process that assures a long lifetime of \ce{CH4} and CO \citep{Hu2012Photo}. In the thermosphere, \ce{H2} becomes increasingly photodissociated, and H becomes ionized. In H-He-dominated atmospheres, most of the absorbed energy from the incident XUV flux feeds the photo-dissociation and -ionization of \ce{H2}, and H and He, respectively, but the excess energy heats the atmosphere and drives atmospheric expansion and escape \citep[e.g.,][]{2014Shematovichefficiency,2016Erkaevmassloss,Salz2016,Linsky2025}.

Similar to hydrogen-rich atmospheres, the lifetimes of \ce{CH4} and CO are long in \ce{CO2}- and \ce{N2}-dominated atmospheres and mostly depend on the availability of OH  \citep{Hu2012Photo}. In reduced \ce{N2}-dominated atmospheres, a rich hydrocarbon chemistry can take place that forms hazes and aerosols, for which CH radicals produced from the photodissociation of \ce{CH4} via far-UV irradiation play an important role \citep[e.g,][]{Trainer2012,Carrasco2022}. This can be observed today at Titan, \citep[e.g.,][]{Lavvas2008}, Pluto and Triton \citep[e.g.,][]{LuspayKuti2025}, and a photochemically produced hydrocarbon haze layer could have also been present on the Archean Earth \citep[e.g.,][]{Pavlov2001}. There, such a layer may have formed for \ce{CH4} and \ce{CO2} abundances of \ce{CH4}/\ce{CO2}$\geq0.1$ \citep[e.g.,][]{Trainer2004,Mak2023}, although with a different chemical composition as on Titan \citep{Mak2023}.

Hazy atmospheres, similar to the Earth's prior to the Great Oxygenation event, as, e.g., recently simulated for TRAPPIST-1e \citep{Mak2024}, may also exist on exoplanets and often serve as example atmospheres of habitats with primitive pre-photosynthesizing life \citep[e.g,][]{EagerNash2024}. However, one has to account for the incident XUV flux and the related thermal stability of these atmospheres. It is unlikely that secondary, Earth-like atmospheres are thermally stable on planets orbiting in the HZ of low-mass M dwarfs such as TRAPPIST-1e \citep{VanLooveren2024Trappist,Scherf2024EEII}. Simulating the photochemistry of specific atmospheres on these planets should, therefore, either include a proper treatment of their thermospheres that also investigates their thermal stability, or the results must be viewed cautiously. If the assumed atmosphere cannot exist because of the high incident XUV flux, neither simulating their photochemistry nor their climates has a practical physical sense.

\section{Summary}
This review highlights that the composition of the atmosphere of a rocky exoplanet is not static in time but evolves throughout the lifetime of the planet through interactions with the planet's interior, stellar irradiation, and, potentially, biological activity.

In order for a rocky planet to retain an atmosphere in the first place, the volatile source flux must exceed the combined loss of volatiles via escape to space and sequestration into the planetary interior. 
There are two main ways how a rocky planet can gain an atmosphere. 
The first is the direct capture of a primordial atmosphere during formation, and the second is through outgassing of volatiles that are initially stored in the interior of the planet. 
These secondary atmospheres are further divided in atmospheres formed by the outgassing of volatiles during the solidification of the magma ocean and volcanic outgassing, which can act over Gyr timescales.

In the first few 100s Myr, the most dominating escape mechanism is hydrodynamic thermal escape. As the planet evolves, the dominating escape process shifts to non-thermal escape processes, such as photochemical escape and stellar-wind escape. The biggest difference between the two escape processes is that thermal escape is in general more efficient at removing lighter species especially for planets that experience XUV fluxes similar to ones of the terrestrial planets in the Solar System , while non-thermal escape processes can also remove heavier species. Rocky planets likely undergo episodic atmospheric replenishment after significant mass-loss events throughout their lifetime.

The volatile mixtures degassed into the atmosphere further undergo significant transformations through selective loss to space, photochemical reactions in the upper atmosphere, and fluid-rock interactions in the lower atmosphere. 
These processes can drastically alter the composition of the volcanic atmosphere compared to the initial composition after outgassing.

The interaction between the interior and the atmosphere continues throughout an active planet's lifetime. On modern Earth the carbonate-silicate cycle in concert with subductive plate tectonics provides a means to stabilize the climate over long time-scales providing long-term habitability. However, even on a planet with a stagnant lid there are means to provide volatile cycling over shorter timescales of order 1 Gyr. With volatile cycling a planet can end up in a moist or runaway greenhouse state (like Venus) or conversely a snowball state via the ice/albedo feedback as seen in Earth's distant past over short periods of time. Clouds play a key role in the radiative budget of a planet, and may even provide a means to stabilize the climate of a planet inside the conservative habitable zone via the cloud-albedo feedback seen in simulations of slowly rotating worlds. In addition to conventional volatile cycling, Nitrogen and Sulfur cycling may also take place. Venus has more than three times Earth's atmospheric nitrogen budget, which is likely tied to its distinct outgassing history from Earth and a long-term lack of plate tectonics (see Section \ref{sec:NCycle} above).

The effect of life on an atmosphere and the search for it via unambiguous biosignatures remains one of the most sought after goals in exoplanetary science. We know that Earth has exhibited biosignature gases such as CH$_4$ since the early Archean and the rise of microscopic life. CH$_4$ could have also played a critical role in resolving the Faint Young Sun Paradox: the fact that the sun was 25\% less luminous $\sim$ 4Gyr ago and was not in a snowball state. Earth's great oxygenation event, driven by cyanobacteria, completely transformed the atmosphere and eventually gave rise to diverse forms of macroscopic life. Yet CH$_4$ and O$_2$ are in fact not unambiguous biosignature gases. For this reason the search of unambiguous biosignature gases remains an active area of research.

The characterization of a large sample of rocky exoplanet atmospheres with current and upcoming telescopes will therefore allow us to deepen our understanding of rocky exoplanets and their interior. 
It is especially important to probe planets across a variety of ages and environments. 
However, interpreting the observations of rocky planet atmospheres necessitates the understanding of the interdependent nature of diverse planetary and atmospheric processes.
This will require not just advancements in instrumentation and observational techniques, but, crucially, the development of numerical models capable of simulating the co-evolution of the atmosphere and interior of rocky planets as a single system.
The development of such models further requires robust knowledge of the physiochemical properties, such as solubility and partitioning coefficients, of key volatiles across a wide range of pressures, temperatures and redox states.  
Furthermore, aerosols remain a major uncertainty in understanding the atmospheres of rocky exoplanets, demanding improvements in microphysical modelling and the acquisition of new laboratory data.
Most importantly, we need to foster strong multidisciplinary collaborations to truly understand and characterize rocky exoplanets.

\backmatter

\bmhead{Acknowledgements}
The authors thank the anonymous reviewer for their insightful comments and the International Space Science Institute and the organizers for hosting and managing the workshop ”The
Geoscience of (Exo)planets: Going beyond habitability”, from which this article originated. We would also like to thank A. Mandell, J. Lustig-Yaeger and R. Hu for useful discussions and comments. MS thanks the Austrian Science Fund (FWF) for the support of the VeReDo research project, grant I6857-N. This project was funded by the European Union (ERC, DIVERSE, 101087755). Views and opinions expressed are however those of the author(s) only and do not necessarily reflect those of the European Union or the European Research Council Executive Agency. Neither the European Union nor the granting authority can be held responsible for them. 

\section*{Declarations}

\begin{itemize}
\item The authors declare that there are no competing interests.
\end{itemize}


\bibliography{library}% common bib file

\end{document}